%% file: draft_arxiv.tex
\documentclass[aps,prd,twocolumn,showpacs,superscriptaddress,groupedaddress]{revtex4-1}  
\usepackage{amssymb}
\usepackage{amsfonts}
\usepackage{amsmath,array}
\usepackage{epsfig}
\usepackage{graphicx}
\usepackage{epstopdf} 
\usepackage{dcolumn}
\usepackage{bm}
\usepackage{lineno}
\usepackage{overpic}
\usepackage{multirow}
\usepackage{color}
\usepackage{bm}
\usepackage[utf8]{inputenc}
\usepackage[perpage,symbol]{footmisc}
\usepackage{subfigure}
\usepackage{enumitem}
\usepackage[colorlinks,
            linkcolor=blue,
            anchorcolor=blue,
            citecolor=blue]{hyperref}
\definecolor{cobalt}{rgb}{0.0, 0.28, 0.67}
\begin{document}

\title{ \Large{Novel method to extract the femtometer structure of strange baryons using the vacuum polarization effect}}

\author{
\begin{small}
  \begin{center}
    \input{authorlist_2022-08-01}
  \end{center}
  \vspace{0.4cm}
\end{small}
}
\noaffiliation{}

\date{\today}
\begin{abstract}
  {One of the fundamental goals of particle physics is to gain microscopic understanding of the strong interaction. Electromagnetic form factors quantify the structure of hadrons in terms of charge and magnetization distributions. While the nucleon structure has been investigated extensively, data on hyperons is still scarce.  It has recently been demonstrated that electron-positron annihilations into hyperon-antihyperon pairs provide a powerful tools to investigate their inner structure. We present a novel method useful for hyperon-antihyperon pairs of different types which exploits the  cross section enhancement  due  to the vacuum polarization effect at the $J/\psi$ resonance. Using the 10 billion $J/\psi$ events collected with the BESIII detector, this allows a thorough determination of the hyperon structure . The result is essentially a precise snapshot of a $\bar\Lambda\Sigma^0$~($\Lambda\bar\Sigma^0$) pair in the making, encoded in the form factor ratio and the phase. Their values are measured to be $R = 0.860\pm0.029({\rm stat.})\pm0.010({\rm syst.})$, $\Delta\Phi_1=(1.011\pm0.094({\rm stat.})\pm0.010({\rm syst.}))~\rm rad$ for $\bar\Lambda\Sigma^0$ and $\Delta\Phi_2=(2.128\pm0.094({\rm stat.})\pm0.010({\rm syst.}))~\rm rad$ for $\Lambda\bar\Sigma^0$, respectively. Furthermore, charge-parity (CP) breaking is investigated for the first time in this reaction and found to be consistent with CP symmetry.}
\end{abstract}
\maketitle

\section{Introduction}
\label{intro}
One distinctive  feature of the strong nuclear interaction and  a prerequisite for our existence is the confinement of nearly  massless quarks into  stable and massive hadrons such as protons or neutrons that constitute the  matter we are made of.  A  coherent understanding of the dynamics of the strong interaction, however, remains one  of the most intriguing puzzles of physics. The main challenge is the very nature of confinement: the quarks and gluons cannot be observed as bare particles, but are “dressed” by the strong interaction into quasi-particles, or constituent quarks, that form the bound systems we know as hadrons. The  distribution and motion of quarks inside hadrons is  quantified in terms of,  e.g.,  electromagnetic form factors which  offer an empirical tool to study the strong dynamics. The proton, as the most stable composite particle we know (with a lifetime much longer than the age of the Universe), offers an excellent testing ground  for the strong interaction. The space-like form factors of the proton have been subject of  rigorous studies since 1956, when Hofstadter  introduced the electron scattering techniques~\cite{Hofstadter}. To  this day, new and  surprising features are being discovered~\cite{EPJA51,PRC85,PLB817} and debated~\cite{PLB816,PRD104,Few-body}.

A  common  strategy to achieve a deeper understanding of these features is to investigate the impact of  introducing heavy and unstable quarks into the bound system. The lightest siblings of the proton are the $\Lambda$ and the $\Sigma^0$ hyperons, both consisting of an up-quark ($u$), a down-quark ($d$) and a heavy and unstable strange-quark ($s$), in contrast to the proton with  a $uud$ structure of only light quarks. Since  hyperons are unstable, they cannot be studied with conventional electron scattering techniques. Instead, their timelike form factors can be accessed in electron-positron annihilations with the subsequent production of a hyperon-antihyperon pair, and in hyperon Dalitz decays. The timelike form factors can be seen as “snapshots” of the time evolution of a hyperon-antihyperon pair, carrying  information about the space-like structure. The  space-like and timelike regions are related via dispersion relations. In recent years, the BESIII collaboration has performed pioneering studies of hyperon form factors~\cite{Nuovo40,POS004,EWC192,LambdaLambda,POS550,EWC199,POS121,JPCS1137,Nuovo42,ACP2130,JPCS1435,PLB814,PLB831,PRD106}. In particular, the self-analyzing  hyperon decays can be used  to measure the hyperon polarization, thereby completely determining  the form factors of the $\Lambda$ hyperon~\cite{Lambdaform}.
\begin{figure*}[htp]
  \begin{center}
    \subfigure{
      \label{strong}
      \includegraphics[width=0.3\textwidth]{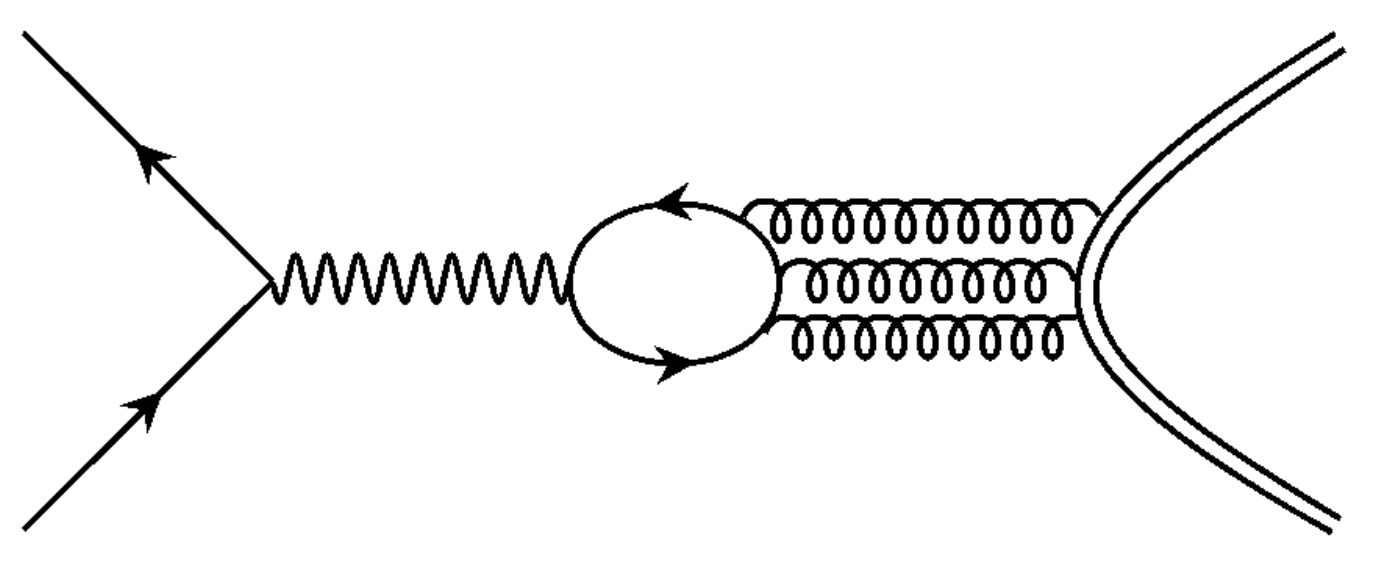}
      \put(-80,45){\footnotesize(a)}\hspace{0.05\textwidth}
    }\subfigure{
      \label{em}
      \includegraphics[width=0.3\textwidth]{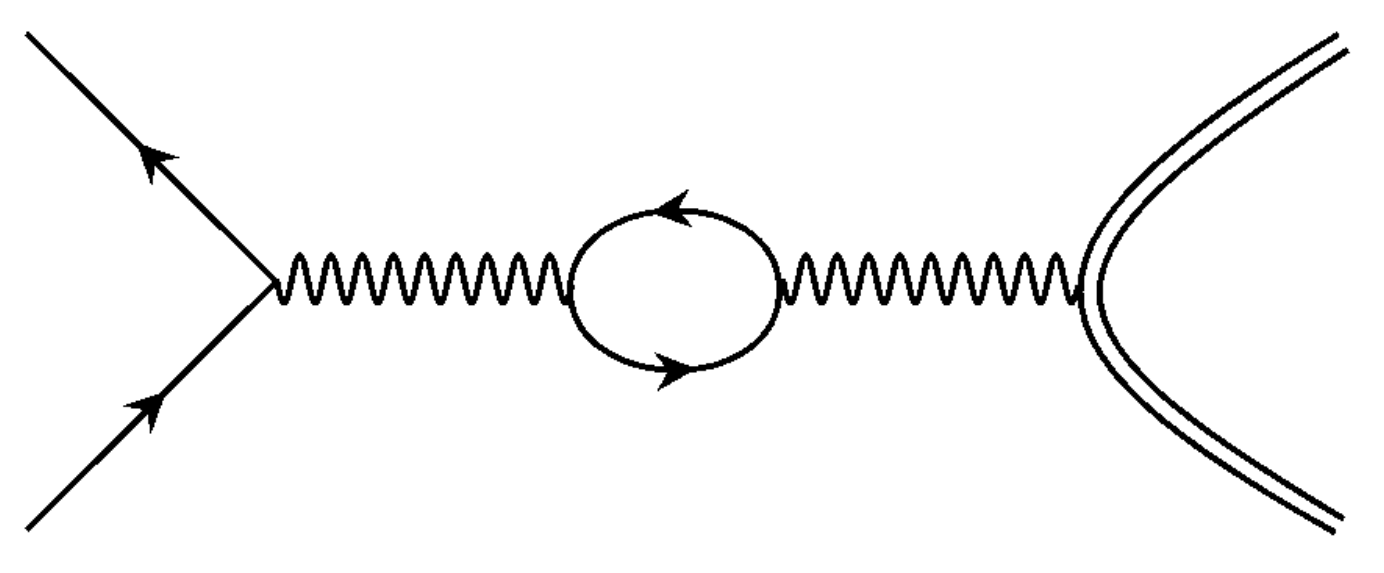}
      \put(-80,45){\footnotesize(b)}
    }\subfigure{
      \label{cont}
      \includegraphics[width=0.3\textwidth]{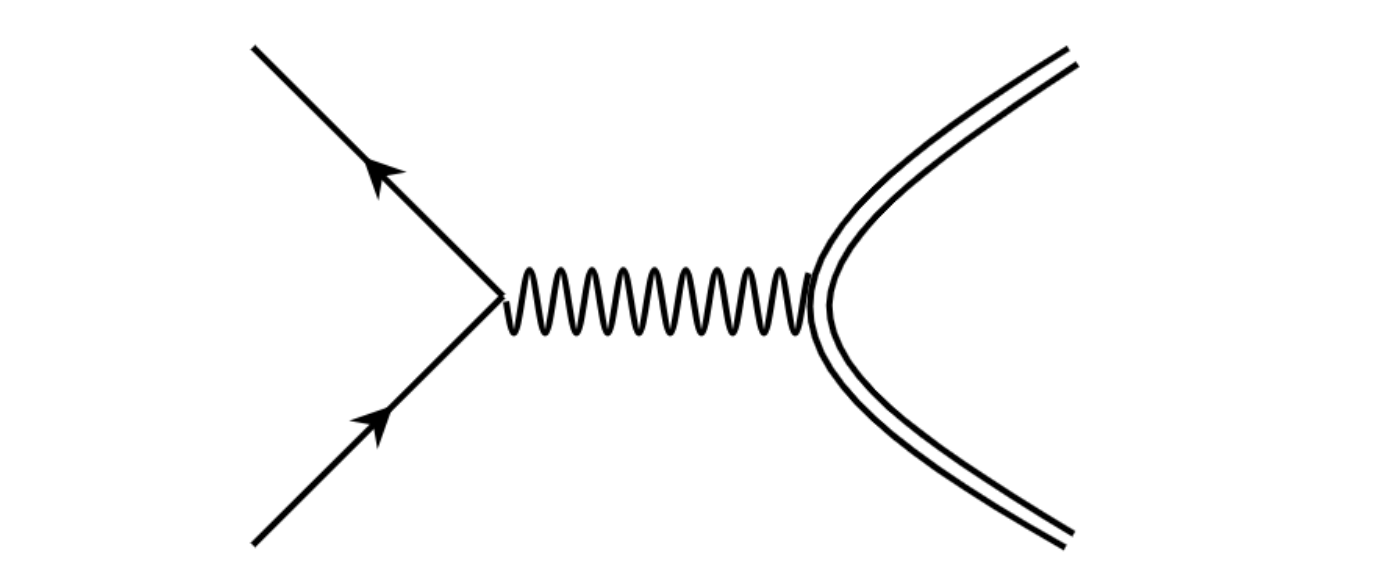}
      \put(-80,45){\footnotesize(c)}
    }
    \caption{The Feynman diagrams for $e^+e^-\rightarrow hadrons$ in the vicinity of the $J/\psi$. (a) strong process with intermediate $J/\psi$ mediated by gluons, (b) electromagnetic process through the vacuum polarization of one virtual photon to a $J/\psi$, (c) continuum process without the $J/\psi$ intermediate state but only one virtual photon.}
    \label{feynman}
  \end{center}
\end{figure*}
However, timelike form factors need to be studied in processes where a  one-photon exchange is the dominating process, as shown in Fig.~\ref{cont}. For a hyperon-antihyperon pair of  the same type, e.g. $\Lambda\bar\Lambda$, this means that the electron-positron annihilation must occur at an energy far from any vector meson resonances that can decay strongly into a hyperon-antihyperon pair. For a pair where the hyperon and the antihyperon are of different type, e.g. $\Lambda\bar\Sigma^{0}$ or $\bar\Lambda\Sigma^{0}$, Ref.~\cite{Simone} proposed the method of comparing the modulus values of the same electromagnetic coupling extracted from the branching ratio of $J/\psi \rightarrow\bar{\Lambda}\Sigma^{0}+c.c.$ and the cross section of the reaction $e^+e^-\rightarrow\bar{\Lambda}\Sigma^{0}+c.c.$ at the $J/\psi$ mass, along with the experimental values of the scaled cross section from BESIII and BaBar data. This suggests that the diagram in  Fig.~\ref{strong} is absent since the process is isospin violating. Hence,  $e^+e^-\rightarrow J/\psi \rightarrow\bar{\Lambda}\Sigma^{0}$ must be a purely electromagnetic process ~\cite{Simone}, as depicted in Fig.~\ref{em}, which has the same final production vertex as Fig.~\ref{cont}. Accordingly, the electric and magnetic form factors of Fig.~\ref{cont} can be extracted from Fig.~\ref{em} by correcting for the  well known vacuum polarization. This demonstrates  that the transition form factor of $\Lambda\bar\Sigma^0$ is accessible both in the low-energy region via Dalitz decay and in  the high-energy region through $e^+e^-$ annihilations~\cite{Perotti:2020smi}, which means we can measure the structure over a much broader energy range than other hyperons. By exploiting quantum entangled pairs of $\Sigma^0~(\bar\Sigma^0)$ and $\bar\Lambda~(\Lambda)$, where $\Sigma^0~(\bar\Sigma^0)$ decays sequentially via intermediate $\bar\Lambda~(\Lambda)$ hyperons, we investigate the reaction $e^+e^-\rightarrow J/\psi \rightarrow\bar{\Lambda}\Sigma^{0}$. We compare this  isospin symmetry breaking and therefore purely electromagnetic transition to the  isospin conserving $J/\psi\rightarrow Y \bar{Y}$ decay, which is an interplay of strong and electromagnetic interactions. With the hadronic vacuum polarization resulting in a significantly enhanced signal, we probe  the same vertex as the one-photon exchange process and attain  the structure at the  $J/\psi$ resonance. At present, the available $(10087\pm44)\times10^6$ $J/\psi$ events produced in $e^+e^-$ annihilations ~\mbox{
\cite{BESIII:2021cxx} }\hspace{0pt}
at  BESIII, almost one order of magnitude larger  than the data sample used in the previous measurement~\cite{LambdaLambda}, offer us a great opportunity to investigate the form factors with the polarized entangled $\Lambda\bar{\Sigma}^0$ pairs. The inclusion of charge-conjugate processes is implied hereafter unless explicitly mentioned otherwise . 
\section{Analysis}

\begin{figure}[htp]
  \begin{center}
      \includegraphics[width=0.45\textwidth]{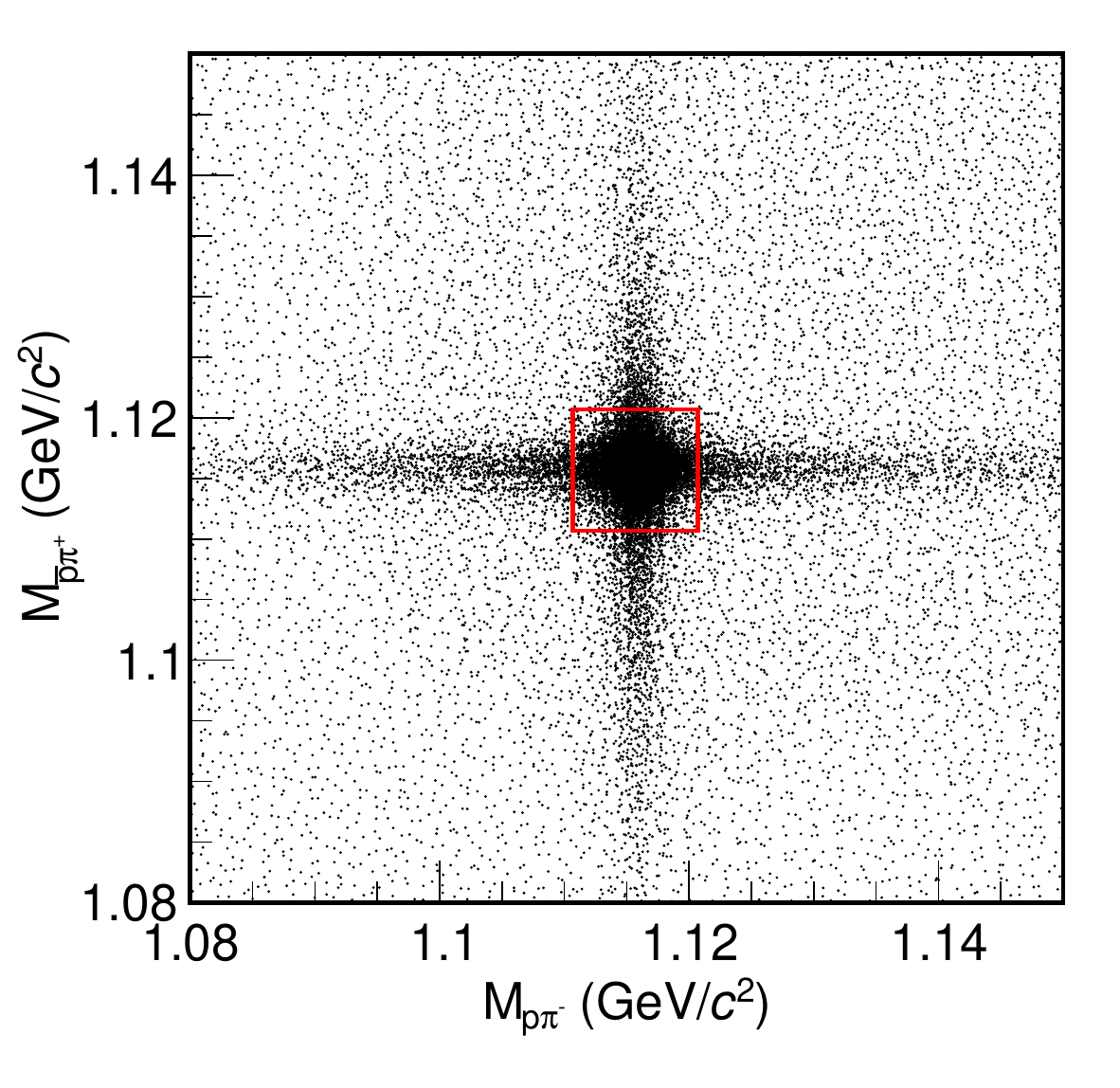}
      \caption{Distribution of $M_{p\pi^-}$ versus $M_{\bar{p}\pi^+}$ of the accepted candidates from data. The red box denotes the signal region. The clusters of $\Lambda$ and $\bar\Lambda$ are clearly observed.}
      \label{scatter_LamLamb}
  \end{center}
\end{figure}

\begin{figure}[htp]
  \begin{center}
      \includegraphics[width=0.5\textwidth]{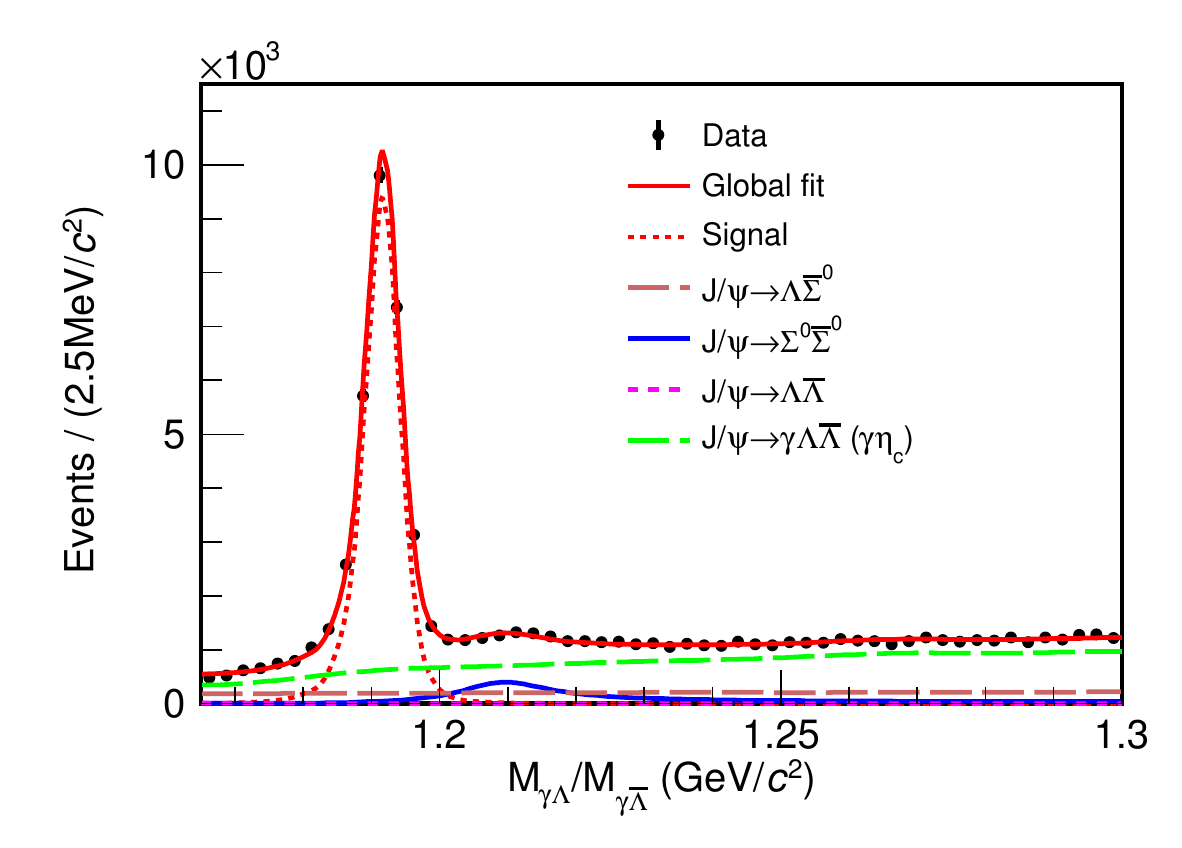}
      \caption{Fit to the $M_{\gamma\Lambda}/M_{\gamma\bar\Lambda}$ distribution of the accepted candidates from $J/\psi$ data. The black points with error bars are data, the red curve is the global fit, the red dotted curve is the $\Sigma^{0}/\bar\Sigma^{0}$ signal, the brown long-dashed curve shows the background from the conjugate channel, the blue curve is from $J/\psi\to\Sigma^0\bar\Sigma^{0}$, the pink dotted curve is from $J/\psi\to\Lambda\bar\Lambda$ and the green long-dashed curve is from $J/\psi\to\gamma\Lambda\bar\Lambda(\gamma\eta_{c})$.}
      \label{combineFits}
  \end{center}
\end{figure}

\begin{figure}[htp]
  \begin{center}
    \includegraphics[width=0.5\textwidth]{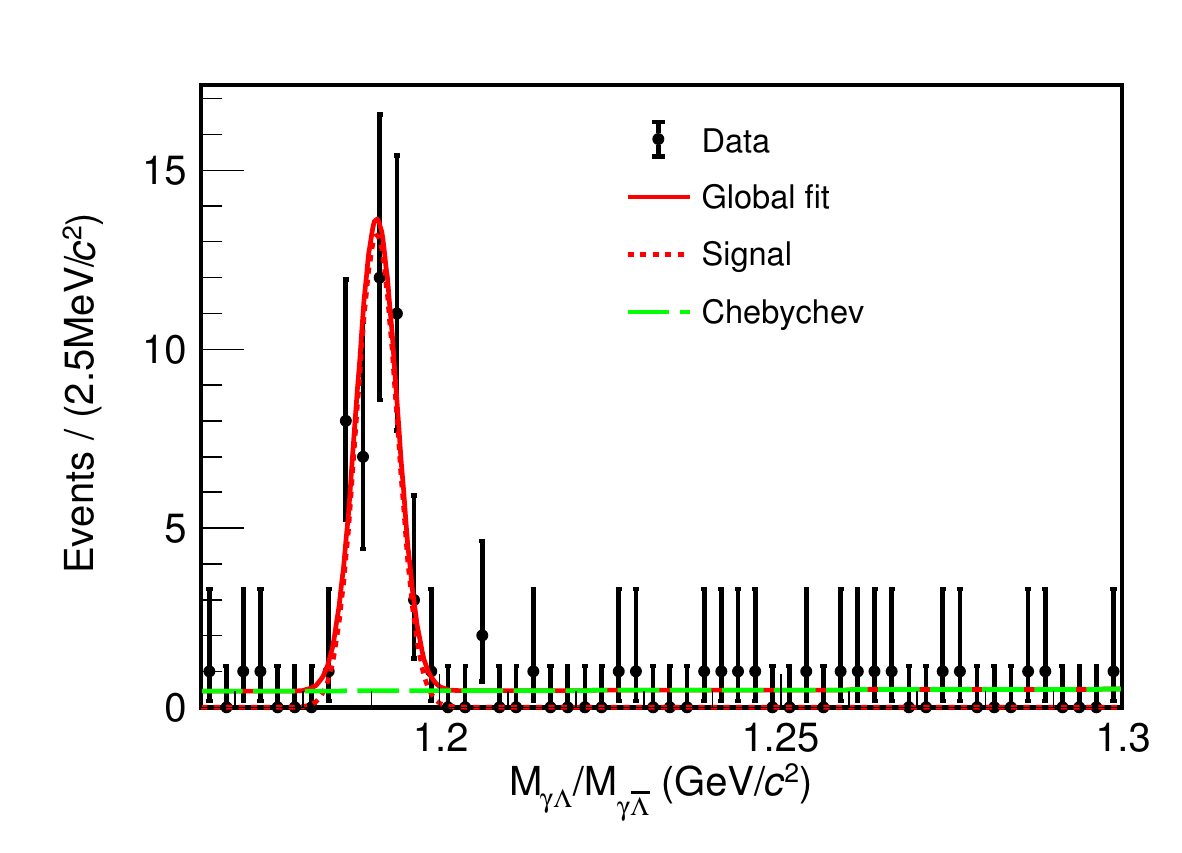}
    \caption{Fit to the $M_{\gamma\Lambda}/M_{\gamma\bar\Lambda}$ distribution from the QED background. The black points with error bars are data, the red curve is the global fit, the red dotted curve is the signal shape described by a Breit-Wigner shape convolved with a Gaussian function  and the green long-dashed curve is from the background described with a first-order Chebychev function.}
    \label{3080_combineFit}
  \end{center}
\end{figure}

The BESIII detector~\cite{Ablikim:2009aa} records symmetric $e^+e^-$ collisions provided by the BEPCII storage ring~\cite{Yu:IPAC2016-TUYA01}, which operates with a peak luminosity of $10^{33}$~cm$^{-2}$s$^{-1}$ in the centre-of-mass energy ($\sqrt{s}$) range from 2.0 to 4.95~GeV. In this cylindrical system, tracks of charged particles in the detector are reconstructed from track-induced signals and the momenta are determined from the track curvature  in the main drift chamber~(MDC) . The flight time of charged particles  is recorded by a plastic scintillator time-of-flight system~(TOF). Showers from photon clusters are reconstructed and energy deposits are  measured in the electromagnetic calorimeter~(EMC). The signal of $e^+e^-\rightarrow J/\psi\rightarrow\bar{\Lambda}(\rightarrow \bar{p}\pi^+)~\Sigma^0(\rightarrow\gamma\Lambda\rightarrow\gamma p\pi^-)$ is extracted from  $(10087\pm44)\times10^6$ $J/\psi$ events~\cite{BESIII:2021cxx} at $\sqrt{s}=3.097$~GeV. The $\Lambda~(\bar\Lambda)$ is reconstructed using $p\pi^-~(\bar{p}\pi^+)$ decays and $\Sigma^0$ from $\gamma\Lambda$ decays. The specific requirements of event reconstruction and selection criteria are described in Appendix~\ref{Icut} and~\ref{Fcut}. The resulting signals of $\bar\Lambda$($\Lambda$) and $\Sigma^0(\bar{\Sigma}^0)$ are clearly observed, as shown in Figs.~\ref{scatter_LamLamb} and~\ref{combineFits}.
The possible background events are investigated with an inclusive Monte Carlo (MC) sample generated with all known $J/\psi$ decays. To estimate the number of background events coming directly from the continuum light hadron (QED) process, the same analysis is performed on the data sample at $\sqrt{s}=3.080$~GeV, corresponding to an integrated luminosity of 166.3 pb$^{-1}$~\cite{BESIII:2021cxx}. With an extended unbinned maximum likelihood fit to the $\gamma\Lambda$ ($\gamma\bar{\Lambda}$) invariant mass distribution shown in Fig.~\ref{3080_combineFit}, the final signal yields are determined to be $26260\pm181$ and the QED background are $39\pm7$.  The details of background analysis and fit are described in Appendix~\ref{Bkg}.

Based on the studies  of $e^+e^-\to\mu^+\mu^-$ and $\eta\pi^+\pi^-$ in Ref.~\cite{eemm},  the relative phase between the hadronic vacuum (Fig.~\ref{em}) and the continuum (Fig.~\ref{cont}) processes is zero in case of a purely  electromagnetic decay, and it has a line shape similar to the cross section of the purely electromagnetic process. Consequently,  the ratio of the cross section at the $J/\psi$ peak to that at any  specific energy is the same for different purely electromagnetic processes as  illustrated by both $e^+e^-\to\mu^+\mu^-$ and $\eta\pi^+\pi^-$. With the measured cross sections in Ref.~\cite{eemm}, the corresponding ratios of these two processes are calculated to be $24.20\pm0.81$ and $28.81\pm8.52$, respectively, both  in good agreement with each other. Here, the uncertainties are statistical only since the systematic uncertainties  cancel in the calculation of the ratio. 
We also performed the measurement of the cross sections of $e^+e^-\to \bar{\Lambda}\Sigma^{0}+c.c.$ 
at the $J/\psi$ peak and at 3.08 GeV, determining  the corresponding ratio to be $33.72\pm6.06$. This value is consistent with those from the above processes within the uncertainties, thus providing further  evidence for $J/\psi \rightarrow\bar{\Lambda}\Sigma^{0}+c.c.$ as a purely  electromagnetic decay, which implies a novel way to extract the electromagnetic form factor with the hadronic vacuum polarization at the $J/\psi$ peak. 

\begin{figure}[htp]
  \begin{center}
      \includegraphics[width=0.5\textwidth]{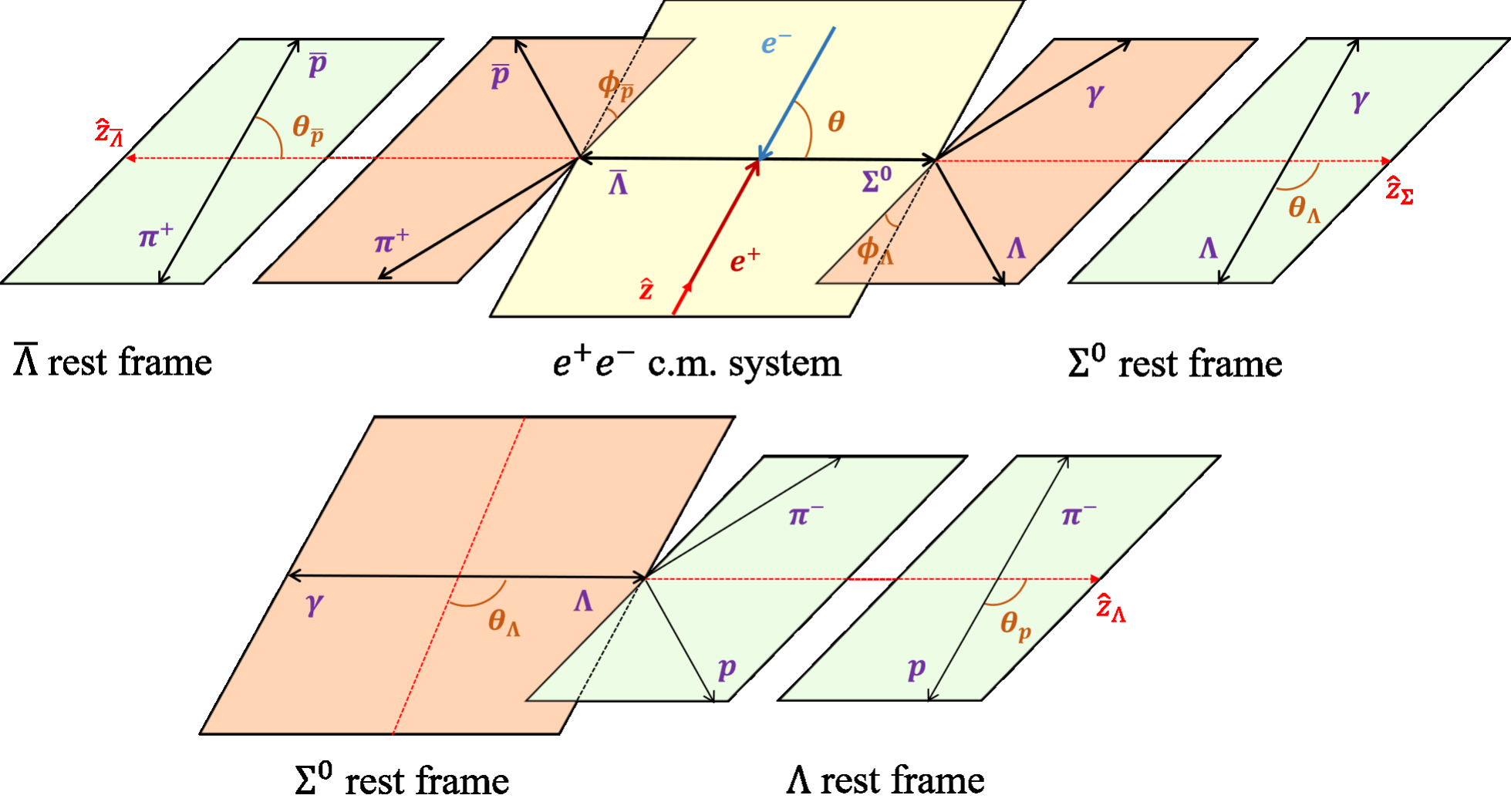}
      \caption{Definition of the helicity angles for $J/\psi\rightarrow\bar{\Lambda}(\rightarrow \bar{p}\pi^+)~\Sigma^0(\rightarrow\gamma\Lambda\rightarrow\gamma p\pi^-)$. The angles $\theta$, $\theta_{\Lambda}$, $\theta_{p}$, $\theta_{\bar{p}}$ are the polar helicity angles of the $\Sigma^0$, $\Lambda$, $p$ and $\bar{p}$ in the $e^+e^-$ center-of-mass system, $\Sigma^0$ rest frame, $\Lambda$ rest frame and $\bar\Lambda$ rest frame, respectively. The angles between different decay or production planes, $\phi_{\Lambda}$ and $\phi_{\bar{p}}$, are the azimuthal helicity angles of the $\Lambda$ and $\bar{p}$ in the $\Sigma^0$ rest frame and $\Lambda$ rest frame, respectively. In the $e^+e^-$ center-of-mass system, the $\bm{\hat{z}}$ is along the $e^+$ momentum direction, and the $\bm{\hat{z}_{\Sigma}}$ is along the $\Sigma^0$ outgoing direction. In the $\Sigma^0$ rest frame, the polar axis is $\bm{\hat{z}_{\Sigma}}$, $\bm{\hat{y}_{\Sigma}}$ is along $\bm{\hat{z}} \times \bm{\hat{z}_{\Sigma}}$ and $\bm{\hat{z}_{\Lambda}}$ is along the $\Lambda$ outgoing direction. In the $\Lambda$ rest frame, the polar axis is $\bm{\hat{z}_{\Lambda}}$, and $\bm{\hat{y}_{\Lambda}}$ is along $\bm{\hat{z}_{\Sigma}} \times \bm{\hat{z}_{\Lambda}}$. In the $\bar\Lambda$ rest frame, the polar axis is $\bm{\hat{z}_{\bar\Lambda}}$, and $\bm{\hat{y}_{\bar\Lambda}}$ is along $\bm{\hat{z}} \times \bm{\hat{z}_{\bar\Lambda}}$.}
      \label{helicity}
    \end{center}
\end{figure}

Since the imaginary part of form factors is non-zero at centre-of-mass energies above  the two-pion threshold ~\cite{Im1-7,Im2-8}, the relative phase $\Delta\Phi$ between the electric and magnetic form factors,  $G_E$ and $G_M$,  is expected to be non-zero . In case of $e^+e^-\rightarrow J/\psi \rightarrow\bar{\Lambda}\Sigma^0$, a non-vanishing $\Delta\Phi$ also demonstrates the polarization of $\Lambda$ and $\bar{\Sigma}^0$  in the direction perpendicular to the production plane. Since the electron mass is negligible in comparison to the $J/\psi$ mass, the initial electron and positron helicities have to be opposite. This implies that the angular distribution and polarization can be described uniquely by only two quantities, the relative phase $\Delta\Phi={\rm arg}(G_E/G_M)$ and the angular distribution parameter $\alpha=\frac{s-4M^2_YR^2}{s+4M^2_YR^2}$~\cite{formfactor}, where $R=|\frac{G_E}{G_M}|$ and $M_Y$ is the mass of the final hyperon. For $\bar{\Lambda}\Sigma^{0}~(\bar{\Lambda}\Sigma^{0})$, $M_Y$ is replaced by $(M_{\Sigma^0}+M_{\Lambda})/2$~\cite{HyperonFFs}.
The probability of extracting the form factors in the production and cascade decays of $e^+e^-\rightarrow J/\psi \rightarrow\bar{\Lambda}(\rightarrow \bar{p}\pi^+)~\Sigma^0(\rightarrow\gamma\Lambda\rightarrow\gamma p\pi^-)$
is described by the six kinematic variables as described in Appendix~\ref{HAmp}, expressed as the helicity angles $\bm{\xi}=(\theta,\theta_{\Lambda},\phi_{\Lambda},\theta_{p},\theta_{\bar{p}},\phi_{\bar{p}})$ shown in Fig.~\ref{helicity}.

Here, we denote the angular distribution parameter, the relative phase and decay asymmetries for $\Sigma^0\to\gamma\Lambda$, $\Lambda\to p\pi^-$,  and $\bar\Lambda\to \bar{p}\pi^+$ as $\alpha_{J/\psi}$, $\Delta\Phi$, $\alpha_{\gamma}$, $\alpha_{\Lambda}$,  and $\alpha_{\bar\Lambda}$, respectively. Subsequently, to extract the form factors, the helicity analysis is performed for $J/\psi\to\bar{\Lambda}\Sigma^{0}+c.c.$ based on the angular distribution as described in detail in Appendix~\ref{HAmp}. Although  $e^+e^-\rightarrow J/\psi \to \Lambda \bar \Sigma^0$ and $e^+e^-\rightarrow J/\psi \to \bar \Lambda \Sigma^0$ are two independent reactions, 
their helicity amplitudes are simply related before and after charge-conjugate and parity transformation. As a result, the relative phases $\Delta\Phi$ of these two decays are expected to satisfy $\Delta\Phi_1+\Delta\Phi_2=\pi$, where $\Delta\Phi_1$ and $\Delta\Phi_2$ denote the relative phases of timelike electric and magnetic form factors for $e^+e^-\rightarrow J/\psi\to\bar{\Lambda}\Sigma^0$ and $e^+e^-\rightarrow J/\psi\to\Lambda\bar{\Sigma}^0$, respectively. Therefore, the simultaneous measurement of $\bar\Lambda\Sigma^0$ and $\Lambda\bar\Sigma^0$ offers the possibility to explore  CP violation by evaluating ${\Delta\Phi}_{\rm CP}=\left\lvert\pi-(\Delta\Phi_1+\Delta\Phi_2)\right\rvert$, which is required to be zero from CP invariance.

\begin{figure*}[htp]
  \begin{center}
    \subfigure{
	    \label{T1_SigmaFit}
	    \includegraphics[width=0.31\textwidth]{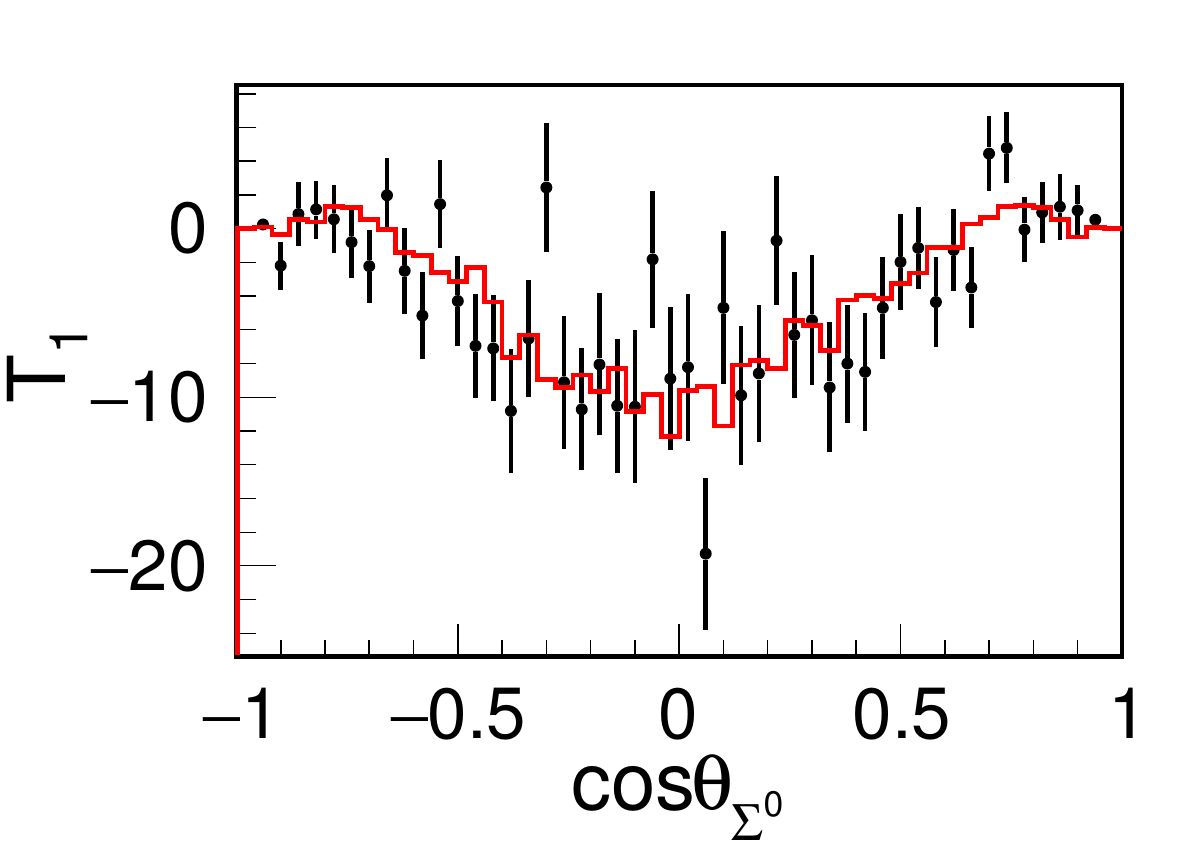}
	    \put(-55,88){\textbf{{\footnotesize(a)}}}
	  }\subfigure{
	    \label{T2_SigmaFit}
	    \includegraphics[width=0.31\textwidth]{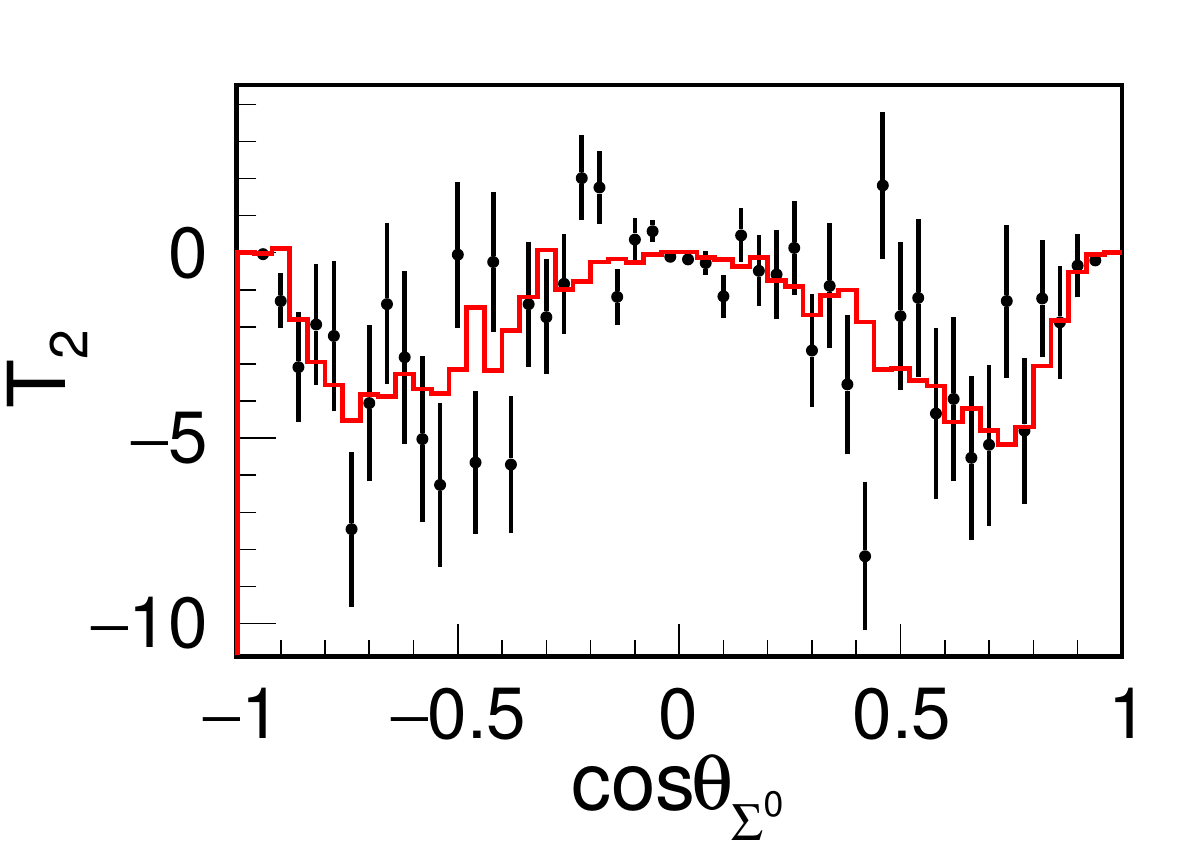}
	    \put(-55,88){\textbf{{\footnotesize(b)}}}
	  }\subfigure{
	    \label{T3_SigmaFit}
	    \includegraphics[width=0.31\textwidth]{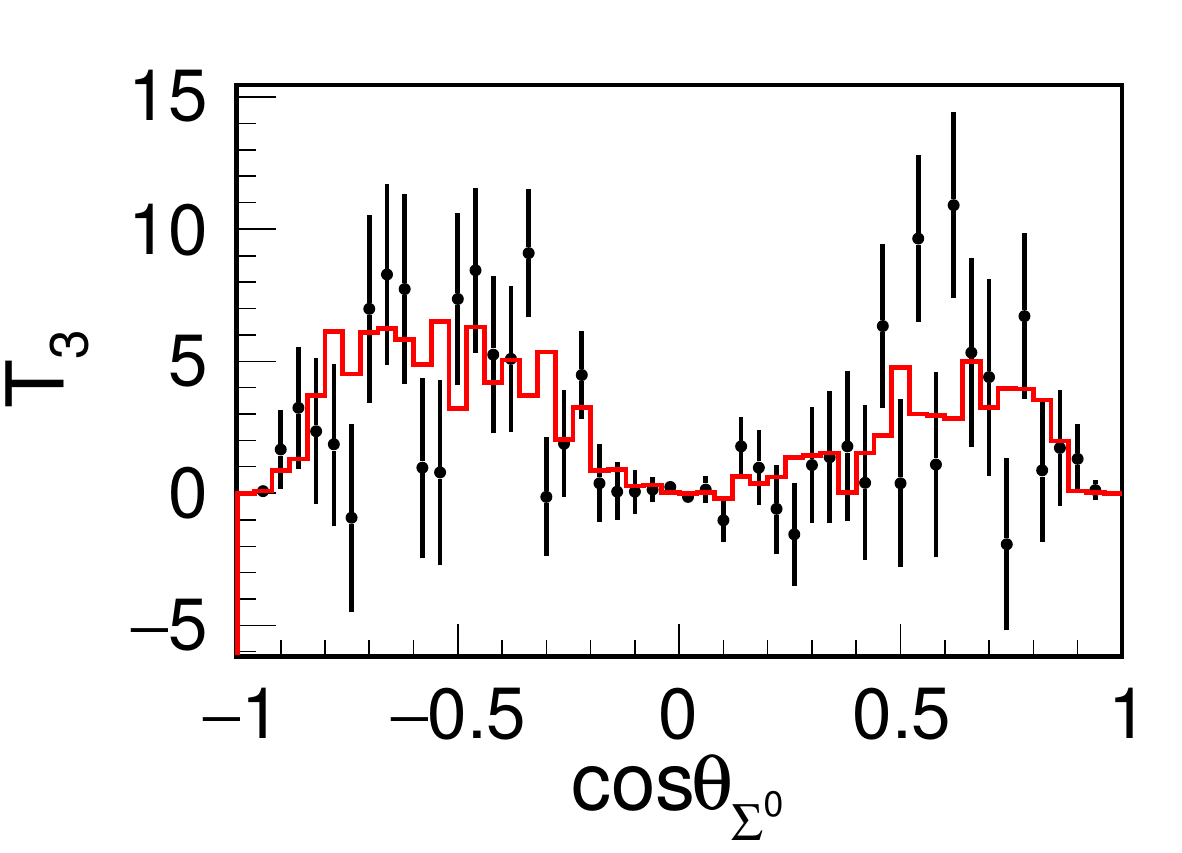}
	    \put(-55,88){\textbf{{\footnotesize(c)}}}
	  }

    \subfigure{
	    \label{T4_SigmaFit}
	    \includegraphics[width=0.31\textwidth]{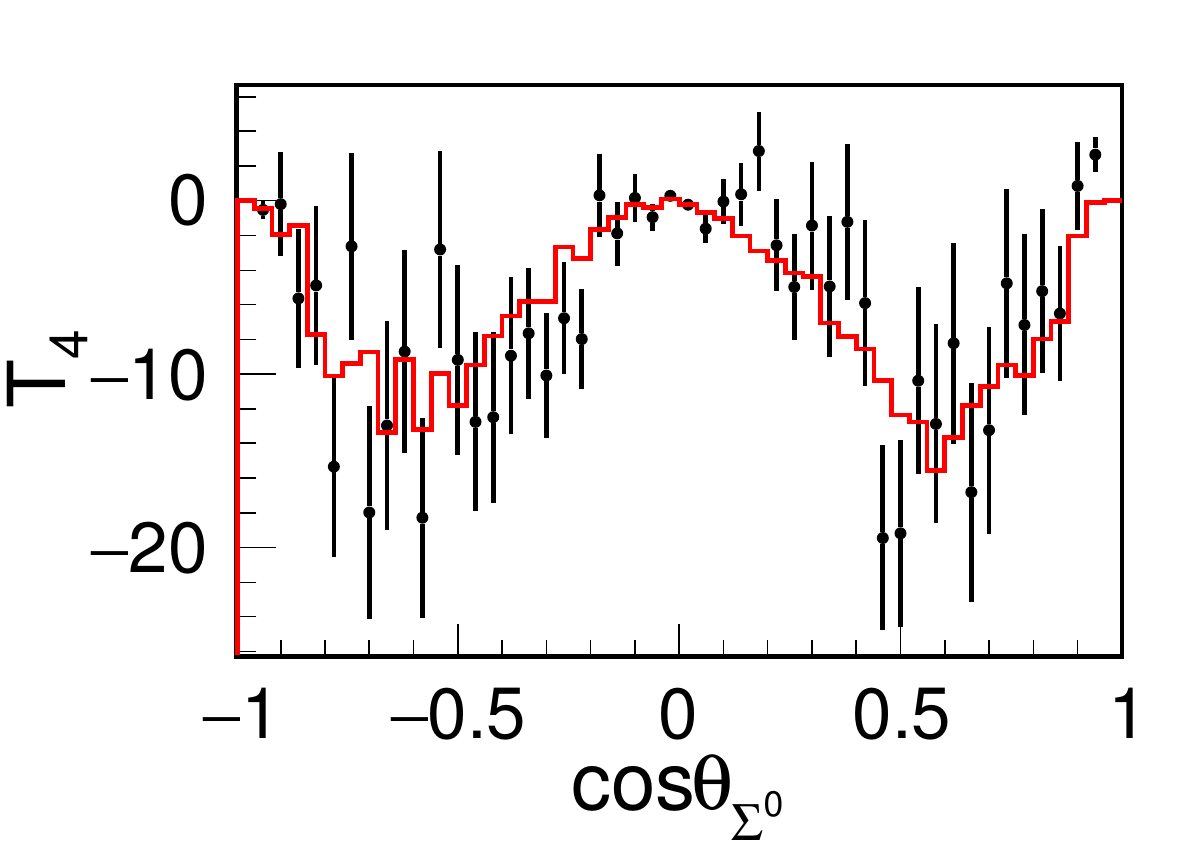}
	    \put(-30,88){\textbf{{\footnotesize(d)}}}
	  }\subfigure{
	    \label{T5_SigmaFit}
	    \includegraphics[width=0.31\textwidth]{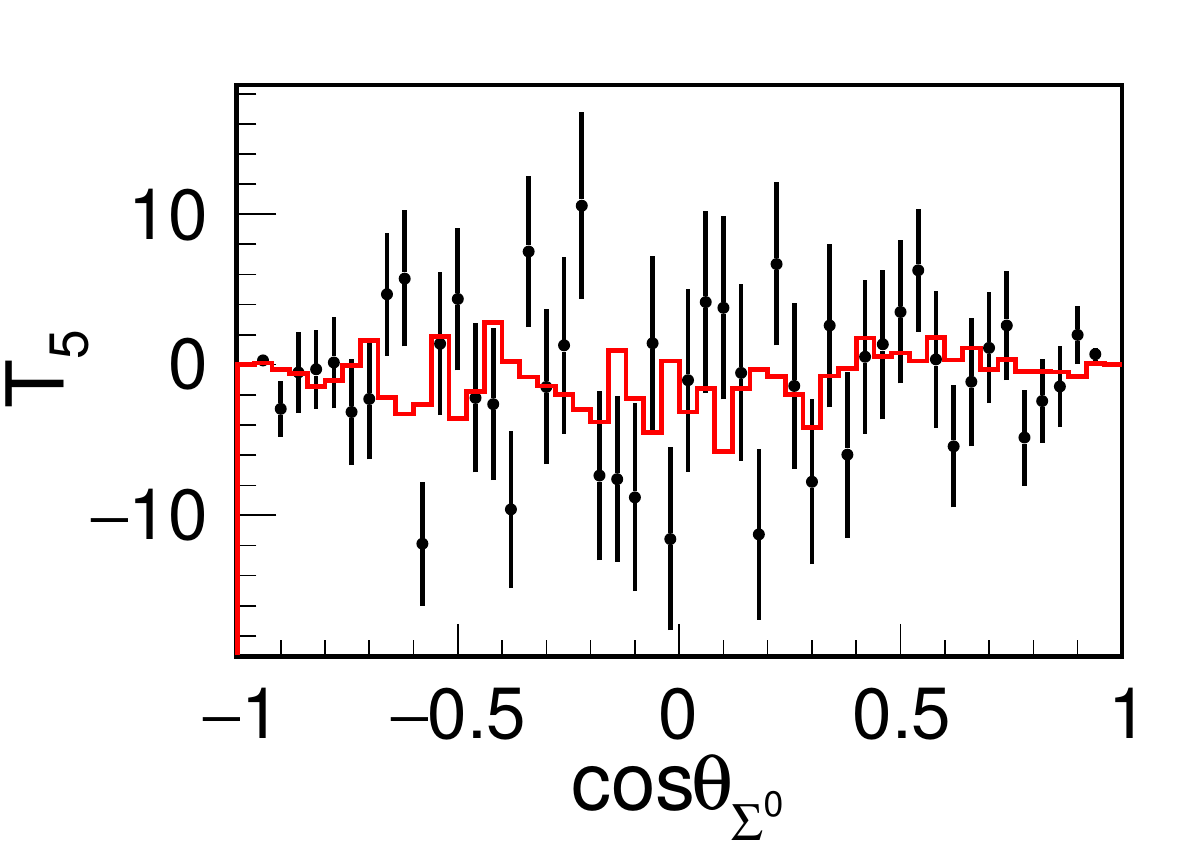}
	    \put(-30,88){\textbf{{\footnotesize(e)}}}
	  }\subfigure{
      \label{Sigma_cos_sigma}
      \includegraphics[width=0.31\textwidth]{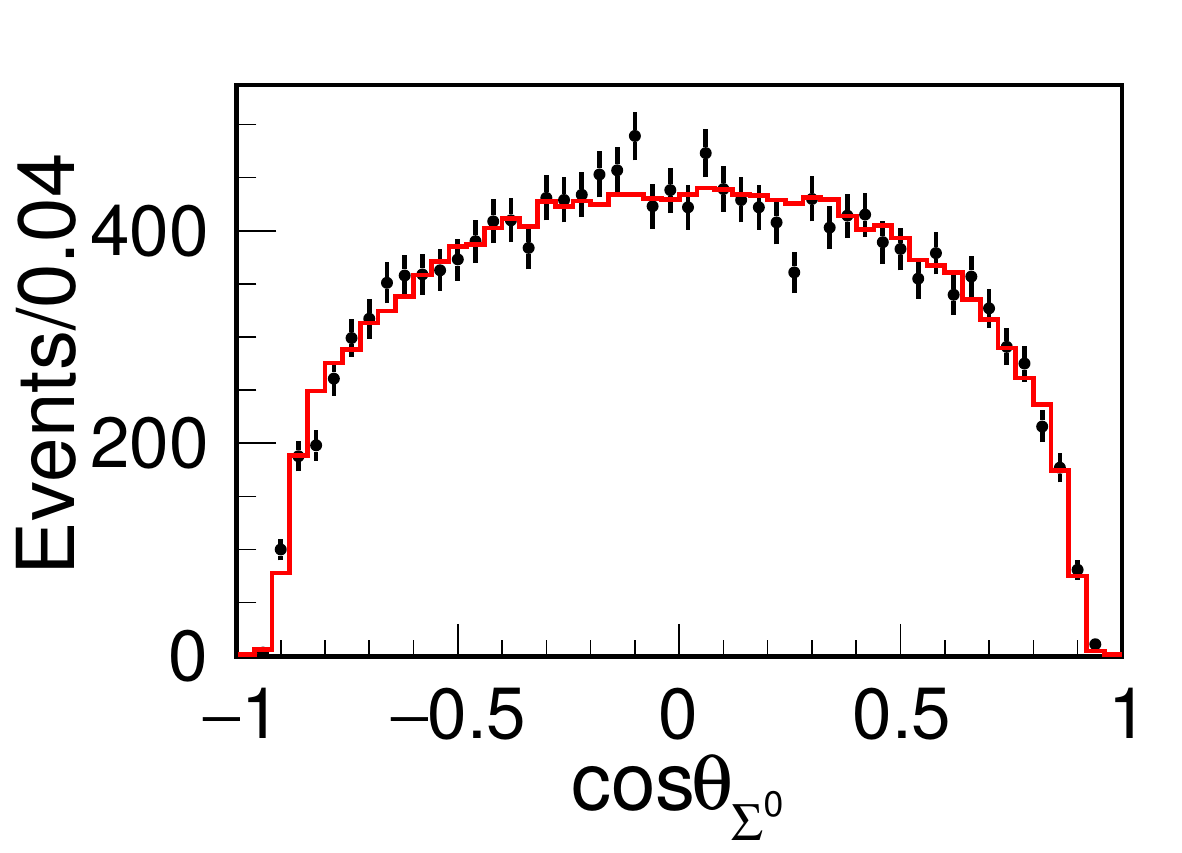}
      \put(-30,88){\textbf{{\footnotesize(f)}}}
    }
	  \caption{Fit results compared with data of $e^+e^-\to J/\psi\to\bar{\Lambda}(\to\bar{p}\pi^+)~\Sigma^0(\to\gamma\Lambda)$ process. (a)-(e) are the moments $T_1$-$T_5$ respectively and (f) is the $\Sigma^0$ angular distribution. The black points with error bars are data and the red curve is the global fit.}  
	  \label{SigmaFitTi}
  \end{center}
\end{figure*}

In the $\Sigma$ mass region, a combined helicity analysis is performed for $J/\psi\to\bar{\Lambda}\Sigma^0$ and $J/\psi\to \Lambda\bar{\Sigma}^0$ and the parameters $\alpha_\Lambda$ and $\alpha_{\bar\Lambda}$ are fixed to be $\alpha_\Lambda=0.7519$ and $\alpha_{\bar\Lambda}=-0.7559$~\cite{BESIII:2022qax} from previous high-precision measurements of  $J/\psi\to \Lambda\bar \Lambda$. Using the average magnitude for both has a negligible affect on fit results.  Due to the electromagnetic part of the decay chain, $\Sigma^0\to\gamma\Lambda$, where the photon polarization is not measured~\cite{gamma}, the $\alpha_\gamma$ is presumed to be 0. The free parameters, including $\alpha_{J/\psi}$ and the relative phase $\Delta\Phi_1$ ($\Delta\Phi_2$) for $e^+e^-\rightarrow J/\psi\to\bar{\Lambda}\Sigma^0~(\Lambda\bar{\Sigma}^0)$, are optimized with an unbinned maximum likelihood fit defined in Appendix~\ref{Fit}. These parameters are measured by incorporating the transverse polarization of $\Sigma^0(\bar\Sigma^0)$ in the joint angular distribution. The global fit is represented by the multidimensional angular distributions shown in Figs.~\ref{SigmaFitTi} and~\ref{SigmaBarFitTi} with a specific fitting technique  as well as systematic uncertainties described in Appendix~\ref{Error}.

\begin{figure*}[htp]
  \begin{center}
    \subfigure{
	    \label{T1_SigmaBarFit}
	    \includegraphics[width=0.31\textwidth]{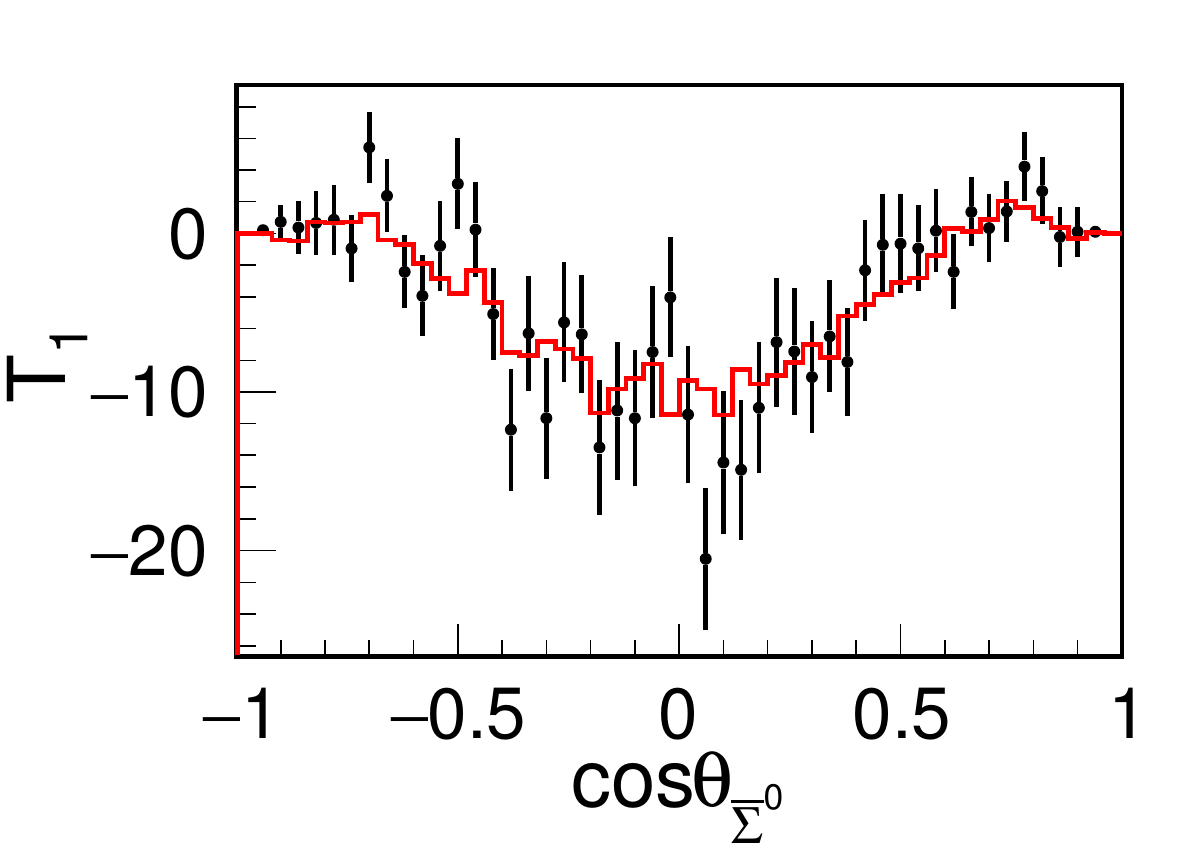}
	    \put(-65,88){\textbf{{\footnotesize(a)}}}
	  }\subfigure{
	    \label{T2_SigmaBarFit}
	    \includegraphics[width=0.31\textwidth]{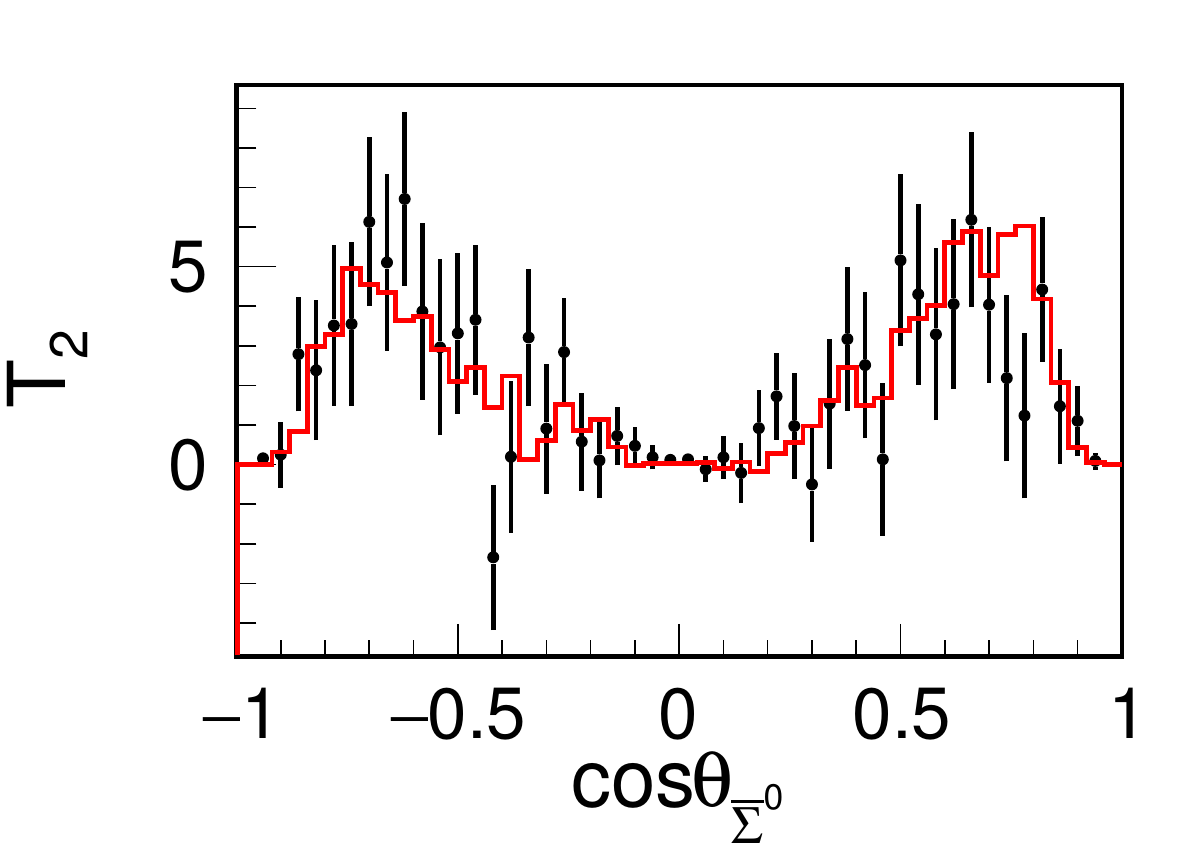}
	    \put(-65,88){\textbf{{\footnotesize(b)}}}
	  }\subfigure{
	    \label{T3_SigmaBarFit}
	    \includegraphics[width=0.31\textwidth]{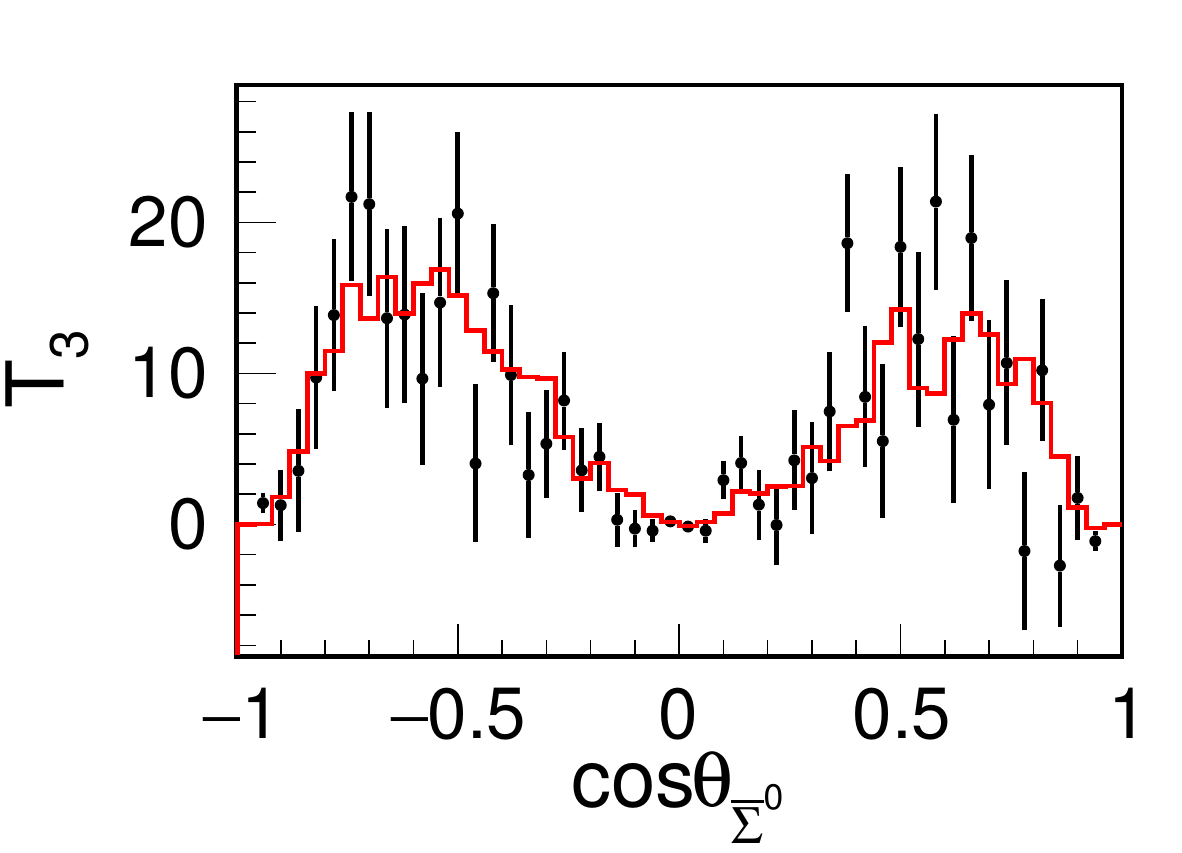}
	    \put(-65,88){\textbf{{\footnotesize(c)}}}
	  }

    \subfigure{
	    \label{T4_SigmaBarFit}
	    \includegraphics[width=0.31\textwidth]{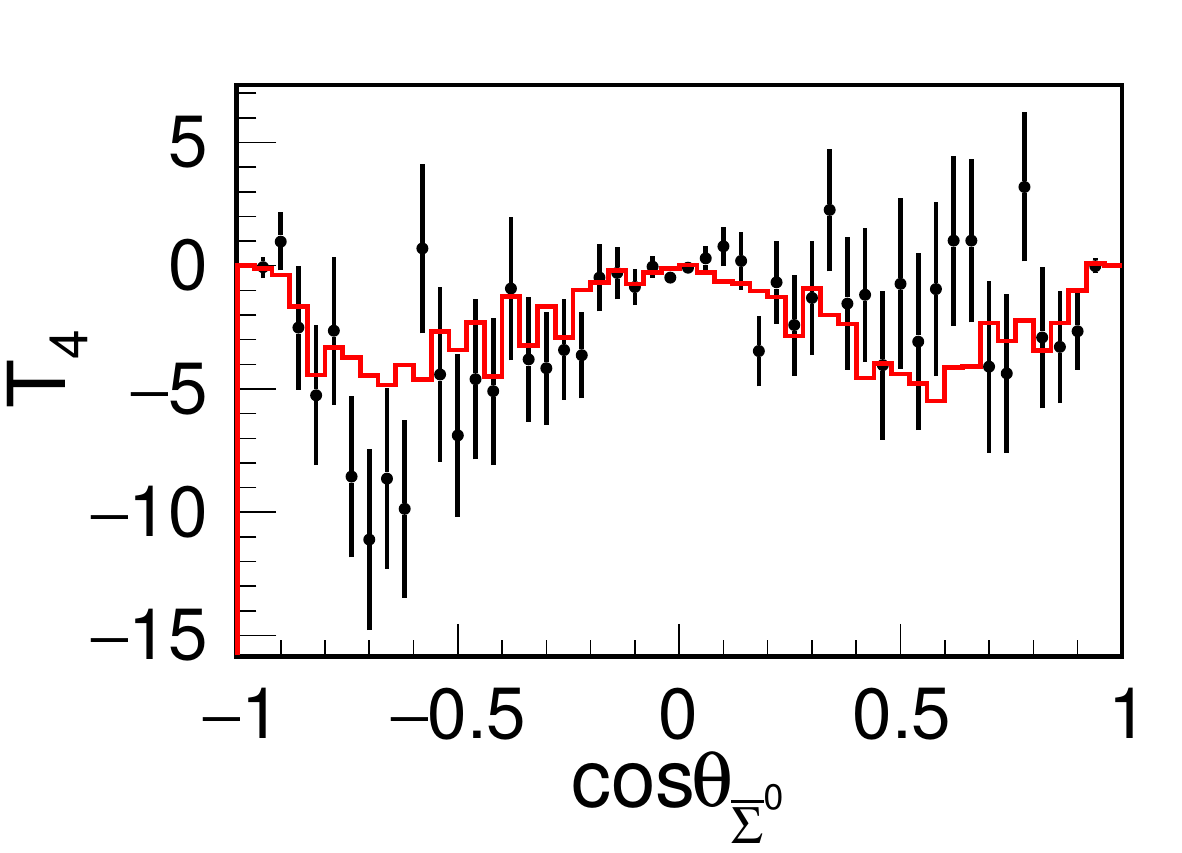}
	    \put(-110,88){\textbf{{\footnotesize(d)}}}
	  }\subfigure{
	    \label{T5_SigmaBarFit}
	    \includegraphics[width=0.31\textwidth]{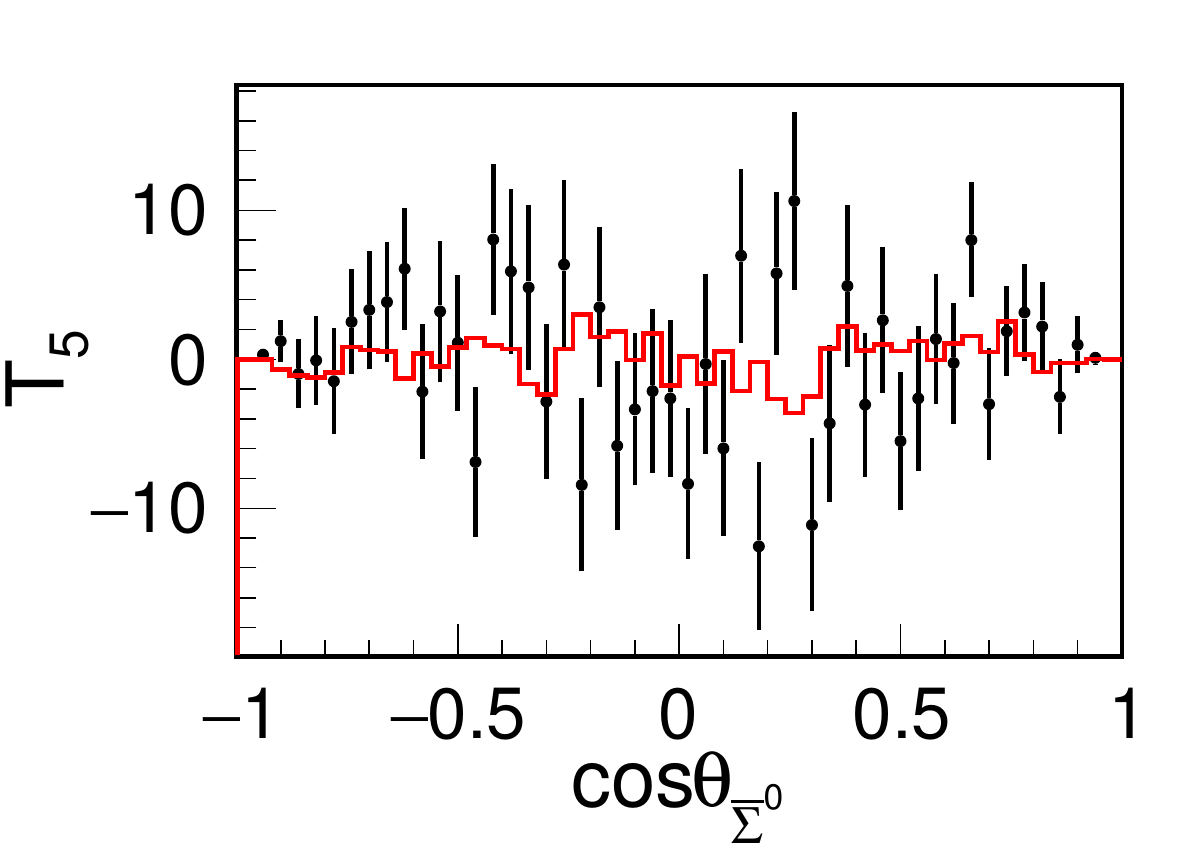}
	    \put(-110,88){\textbf{{\footnotesize(e)}}}
	  }\subfigure{
      \label{Sigmab_cos_sigma}
      \includegraphics[width=0.31\textwidth]{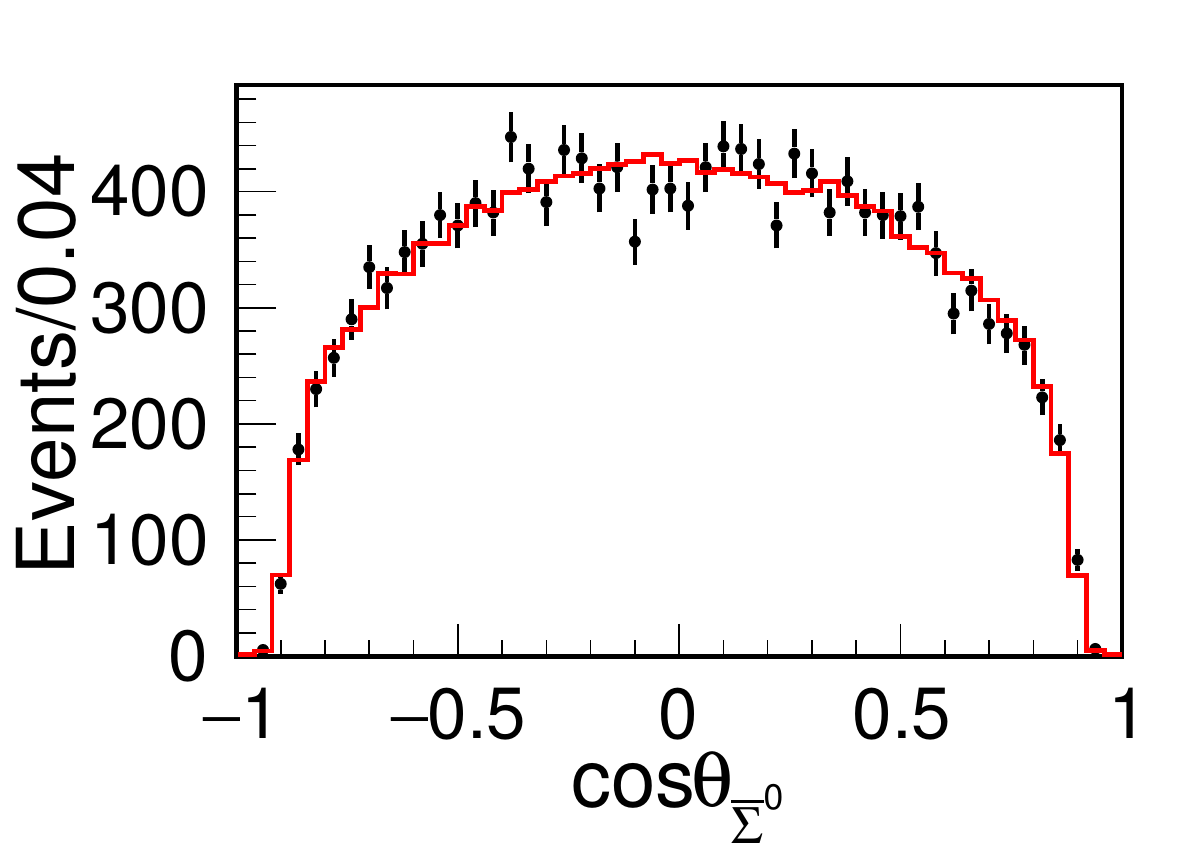}
      \put(-110,88){\textbf{{\footnotesize(f)}}}
    }
	  \caption{Fit results compared with data of $e^+e^-\to J/\psi\to\Lambda(\to p\pi^-)~\bar{\Sigma}^0(\to\gamma\bar\Lambda)$ process. (a)-(e) is the moments $T_1$-$T_5$ respectively and (f) is the $\bar\Sigma^0$ angular distribution. The black points with error bars are data and the red curve is the global fit}  
	  \label{SigmaBarFitTi}
  \end{center}
\end{figure*}
\begin{figure*}[htp]
  \begin{center}
    \subfigure{
      \label{Py}
      \includegraphics[width=0.5\textwidth]{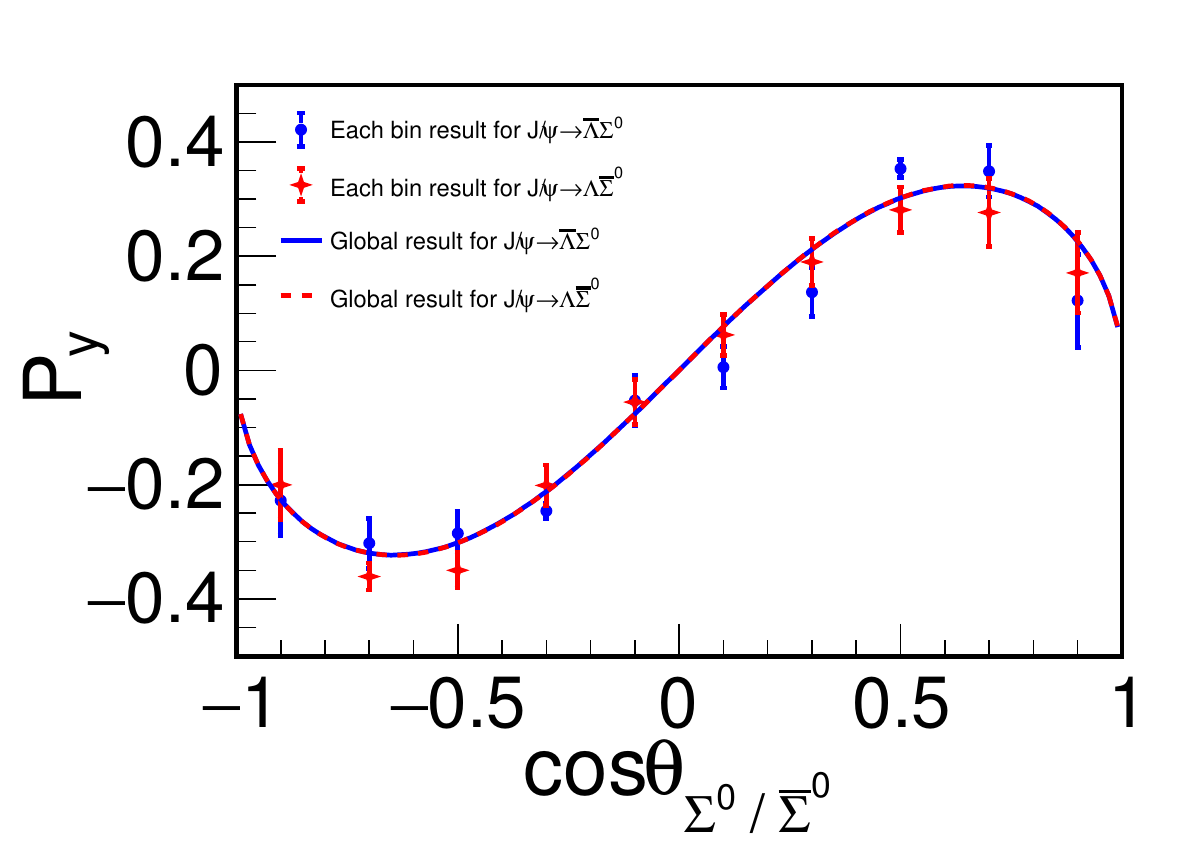}
      \put(-30,50){\footnotesize(a)}
    }\subfigure{
      \label{Cxz}
      \includegraphics[width=0.5\textwidth]{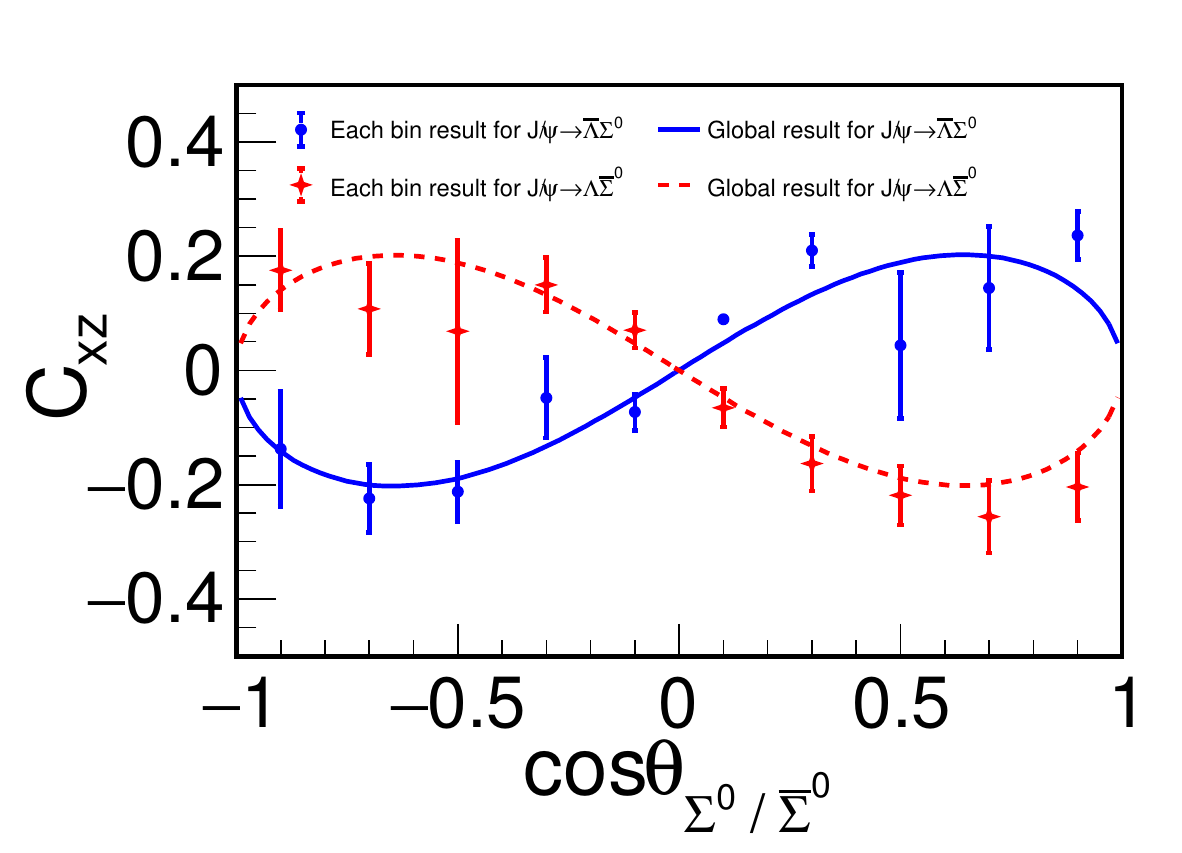}
      \put(-30,50){\footnotesize(b)}
    }
    \caption{Polarization $P_y$ and spin correlations $C_{xz}$ in $e^+e^-\rightarrow J/\psi\to\bar\Lambda\Sigma^0(\Lambda\bar{\Sigma}^0)$. The {points} with error bars, blue solid dot for $J/\psi\to\bar\Lambda\Sigma^0$ and red open double diamond for $J/\psi\to\Lambda\bar{\Sigma}^0$, are extracted in each $\cos\theta_{\Sigma^0}$($\cos\theta_{\bar{\Sigma}^0}$) bin, and the blue solid {curves} denote the global expected dependence on $\cos\theta_{\Sigma^0}$($\cos\theta_{\bar{\Sigma}^0}$ for the red dotted curve).}
    \label{Py_Cxz}
  \end{center}
\end{figure*}

\section{Results}
From the global fit, a  prominent polarization and strong correlation of the relative phase between the two processes are observed,  characterized by $P_{y}$ elucidating the spin transverse polarization and $C_{xz}$ representing the particular relationship between $\Delta\Phi_1$ and $\Delta\Phi_2$. Their strong dependence on the $\Sigma^0~(\bar\Sigma^0)$ direction angle $\theta$, defined in Appendix~\ref{HAmp}, is seen in Fig.~\ref{Py_Cxz}. To illustrate the fit quality, the fit results in each cos$\theta_{\Sigma^0 / \bar \Sigma^0}$ bin are also shown using points with error bars  in Fig.~\ref{Py_Cxz}. Apart from the difference caused by the fluctuations from the complex background channels, the points of each bin are consistent with the globally fitted curves.

The fit yields  $\alpha_{J/\psi}=0.418\pm0.028({\rm stat.})\pm0.010({\rm syst.})$, $\Delta\Phi_1=(1.011\pm0.094({\rm stat.})\pm0.010({\rm syst.}))~\rm rad$,  and $\Delta\Phi_2=(2.128\pm0.094({\rm stat.})\pm0.010({\rm syst.}))~\rm rad$. The ratio  $R=|\frac{G_E}{G_M}|=\frac{\sqrt{s}}{2M_Y} \sqrt{\frac{1-\alpha}{1+\alpha}}$ is determined to be $0.860\pm0.029({\rm stat.})\pm0.010({\rm syst.})$, giving the ratio and relative phase of the electric and magnetic form factors $G_E$ and $G_M$ for $e^+e^-\rightarrow J/\psi \rightarrow\bar\Lambda\Sigma^0~(\Lambda\bar{\Sigma}^0)$ at $\sqrt s=3.097$~GeV, with clear transverse spin polarizations of the $\Lambda$ and $\bar{\Sigma}^0$ observed. The sum of these two relative phases, {$\Delta\Phi_1+\Delta\Phi_2=(3.139\pm0.133({\rm stat.})\pm0.014({\rm syst.}))~\rm rad$}, is in good agreement with the expected value of $\pi$. ${\Delta\Phi}_{\rm CP}=\left\lvert\pi-(\Delta\Phi_1+\Delta\Phi_2)\right\rvert$ is calculated to be $0.003\pm0.133({\rm stat.})\pm0.014({\rm syst.})$, which is consistent with zero and indicates no evident direct CP violation in the decays of $J/\psi\to\bar\Lambda\Sigma^0$ and $J/\psi\to\Lambda\bar\Sigma^0$. It is the first time that the timelike structure for $e^+e^-\rightarrow\bar\Lambda\Sigma^0+c.c.$ is  extracted at $\sqrt{s}=3.097$ GeV with high precision by using the hadronic vacuum polarization enhancement at the $J/\psi$. 
In addition, unlike $e^+e^-$ annihilation into hyperon anti-hyperon pairs, $\Lambda$ and $\bar{\Sigma}^0$ are not charge conjugates of each other which enables us to explore direct CP violation by comparison of polarizations from 
 both $e^+e^-\rightarrow J/\psi \rightarrow\Lambda\bar{\Sigma}^0$  and $e^+e^-\rightarrow J/\psi \rightarrow\bar\Lambda\Sigma^0$.  While currently statistically limited, it provides a novel way to search for possible new sources of CP violation. In the future, the BESIII experiment may provide even higher sensitivity  to  direct CP violation~\cite{whitebook}, with further improvement expected  from the next generation experiments, e.g., 
the next-generation tau-charm physics facility~\cite{ref1} and PANDA~\cite{ref2}.

\bigskip
\acknowledgments
The BESIII collaboration thanks the staff of BEPCII and the IHEP computing center for their strong support. This work is supported in part by National Key R\&D Program of China under Contracts Nos. 2020YFA0406300, 2020YFA0406400; National Natural Science Foundation of China (NSFC) under Contracts Nos. 11635010, 11735014, 11835012, 11875115, 11935015, 11935016, 11935018, 11961141012, 12022510, 12025502, 12035009, 12035013, 12075250, 12165022, 12192260, 12192261, 12192262, 12192263, 12192264, 12192265, 12225509; the Chinese Academy of Sciences (CAS) Large-Scale Scientific Facility Program; Joint Large-Scale Scientific Facility Funds of the NSFC and CAS under Contract No. U1832207; the CAS Center for Excellence in Particle Physics (CCEPP); 100 Talents Program of CAS; The Institute of Nuclear and Particle Physics (INPAC) and Shanghai Key Laboratory for Particle Physics and Cosmology; ERC under Contract No. 758462; European Union's Horizon 2020 research and innovation programme under Marie Sklodowska-Curie grant agreement under Contract No. 894790; German Research Foundation DFG under Contracts Nos. 443159800, 455635585, Collaborative Research Center CRC 1044, FOR5327, GRK 2149; Istituto Nazionale di Fisica Nucleare, Italy; Ministry of Development of Turkey under Contract No. DPT2006K-120470; National Science and Technology fund; National Science Research and Innovation Fund (NSRF) via the Program Management Unit for Human Resources \& Institutional Development, Research and Innovation under Contract No. B16F640076; Olle Engkvist Foundation under Contract No. 200-0605; STFC (United Kingdom); Suranaree University of Technology (SUT), Thailand Science Research and Innovation (TSRI), and National Science Research and Innovation Fund (NSRF) under Contract No. 160355; Polish National Science Centre under Contract 2019/35/O/ST2/02907; The Royal Society, UK under Contracts Nos. DH140054, DH160214; The Swedish Research Council; U. S. Department of Energy under Contract No. DE-FG02-05ER41374.


\appendix{ \section{}}
\subsection{ {Monte Carlo simulation} \label{MC}}
The optimization of the event selection criteria and the estimation of physics background as well as the determination of efficiency are performed using MC simulated samples. The {\sc geant4}-based~\cite{geant4} MC package includes the geometric description of the BESIII detector and the detector response. The inclusive MC sample includes both the production of the $J/\psi$ resonance and the continuum processes incorporated in {\sc kkmc}~\cite{ref:kkmc}. All particle decays are modelled with {\sc evtgen}~\cite{ref:evtgen} using branching fractions either taken from the Particle Data Group~(PDG)~\cite{PDG}, when available, or otherwise estimated with {\sc lundcharm}~\cite{ref:lundcharm}. For the signal $J/\psi\to\bar\Lambda\Sigma^0+c.c.$, the MC samples are produced using the angular distribution formula shown in Appendix~\ref{HAmp}. For the determination of the cross section, the generator {\sc ConExc}~\cite{conexc} was used . For the background channels $J/\psi\rightarrow\Sigma\bar{\Sigma}^0$, $J/\psi\rightarrow\Lambda\bar\Lambda$, the exclusive MC samples were generated in accordance with their decay amplitudes~\cite{Sigma0,LambdaLambda}.

\subsection{Initial Selection Criteria\label{Icut}}
Candidates for $J/\psi\rightarrow\bar{\Lambda}(\rightarrow \bar{p}\pi^+)~\Sigma^0(\rightarrow\gamma\Lambda\rightarrow\gamma p\pi^-)$ are required to have four charged tracks with net zero charge and at least one photon.

Charged tracks are selected in the MDC within $\pm$20 cm of the interaction point in the beam direction and within 10 cm in the plane perpendicular to the beam. The polar angles of these tracks are required to be within the MDC fiducial volume, $|\cos\theta|<0.93$, where $\theta$ is defined with respect to the $z$-axis, which is the symmetry axis of the MDC. No particle identification is performed to maintain high efficiency. 

To reconstruct the decays $\Lambda\to p\pi^-$ and $\bar{\Lambda}\to \bar{p}\pi^+$, we loop over all the combinations of positive and negative charged track pairs and require that at least one ($p\pi^-$)($\bar{p}\pi^+$) track hypothesis successfully passes the vertex finding algorithm~\cite{VertexFit} of $\Lambda$ and $\bar \Lambda$. If more than one accepted combination satisfies  the vertex fit requirement, the one  with the minimum value of $\sqrt{(M_{p\pi^-} - M_\Lambda)^2 + (M_{\bar{p}\pi^+} - M_{\Lambda})^2}$ is chosen, where $M_{p\pi^-} (M_{\bar{p}\pi^+})$ is the $p\pi^- (\bar{p}\pi^+)$ invariant mass and $M_\Lambda$ is the nominal $\Lambda$ mass~\cite{PDG}.

For good photon selection, showers in the EMC identified as photon candidates are required to satisfy fiducial and shower-quality requirements. For the  barrel region, showers must have a minimum energy deposition of 25 MeV with the polar angle of each track satisfying  $|\cos\theta|<0.80$, while those from the end cap region must have at least 50 MeV and the polar angle is required to be $0.86<|\cos\theta|<0.92$. To remove the noise unrelated with the event, the difference between the EMC time and the event start time (TDC) has to fulfill  $0\le {\rm TDC} \le 700$ ns. To suppress showers generated by charged particles, the photon candidate angular separation from the nearest charged track is required to be at least $10^\circ$.

The selected events are subjected to a four-constraint energy momentum conservation kinematic fit (4C fit) with the hypothesis of $\gamma \Lambda\bar{\Lambda}$.
The kinematic fit adjusts the reconstructed particle energy and momentum
within the measured errors so as to satisfy
energy and momentum conservation for the given
event hypothesis. 
This improves resolution and reduces background. 
When there are multiple photon candidates in an event, the combination with
the smallest $\chi^2_{4\rm C}$ is retained. The kinematic fit is very powerful to suppress background events with multiple photon candidates in the final states, e.g., $J/\psi\to\Sigma^0\bar\Sigma^0$ and $J/\psi\to\Lambda\bar\Sigma^0\pi^0$.

\subsection{Final Selection Criteria\label{Fcut}}

After the initial selection, the scatter plot of $M_{p\pi^-}$ versus $M_{\bar{p}\pi^+}$ of the accepted candidates is shown in Fig.~\ref{scatter_LamLamb}, where the clear cluster corresponds to the decays of $\Lambda\to p\pi^-$ and $\bar{\Lambda}\to\bar{p}\pi^+$. The $\Lambda$ and $\bar{\Lambda}$ signal candidates are selected by requiring $|M_{p\pi^-}-M_{\Lambda}|<5$ MeV/$c^2$ and $|M_{\bar{p}\pi^+}-M_{\Lambda}|<5$ MeV/$c^2$. To further suppress backgrounds and improve the mass resolution, the 4C kinematic fit must satisfy $\chi^2_{\rm 4C}<30$. In addition, $M_{\gamma\bar{\Lambda}}>1.135$ GeV/{$c^2$} and $M_{\gamma\Lambda}>1.135$ GeV/{$c^2$} are required in the further analysis for $J/\psi\to\bar\Lambda\Sigma^0$ and $J/\psi\to\Lambda\bar{\Sigma}^{0}$, respectively, which has a pronounced effect on suppressing the background events from $J/\psi\to\Lambda\bar{\Lambda}$. After applying  the above requirements, the invariant mass spectrum of $\gamma\Lambda~(\gamma\bar{\Lambda})$ is shown in Fig.~\ref{combineFits}, where the prominent peak of {$\Sigma^0~(\bar{\Sigma}^0)$} is clearly observed.

\subsection{Background Analysis\label{Bkg}}
Possible background sources are investigated with an inclusive MC sample of 10 billion $J/\psi$ decays. Using the same selection criteria, with the help of a generic event type analysis tool~\cite{TopoAna}, the surviving background events mainly originate from $J/\psi\to\Sigma^0\bar\Sigma^0$, $J/\psi\to\Lambda\bar{\Lambda}$ and $J/\psi\to\gamma\Lambda\bar{\Lambda}$~(including a resonant contribution from $\gamma\eta_c$), but none of these produce an evident peak in the $\Sigma^0$ mass region. The exclusive MC samples of these background channels are generated with the corresponding helicity amplitudes and their contributions are shown in Fig.~\ref{combineFits}. 
To estimate the number of background events coming
directly from the $e^+e^-$ annihilation, the same analysis
is performed on data taken at $\sqrt s=$ 3.080 GeV, where the number of background events, $39\pm7$ is also extracted by fitting the $\gamma\Lambda$ (or $\gamma\bar{\Lambda}$) mass spectrum as shown in Fig.~\ref{3080_combineFit}. 
The background events are then normalized to the $J/\psi$ data after taking into account the luminosities and energy-dependent cross sections of continuum processes ~\cite{QED}, with  the scaling factor calculated as
  \begin{eqnarray} \begin{aligned}
 f=\frac{\mathcal{L}_{J / \psi}}{\mathcal{L}_{\psi(3080)}} \times \frac{s_{\psi(3080)}^{5}}{s_{J/\psi}^{5}} \times \frac{{\epsilon}_{\psi(3080)}}{{\epsilon}_{J/\psi}}.
 \label{fQED}
\end{aligned} \end{eqnarray}
Here,  $\mathcal{L}$, $s$,  and $\epsilon$ are the integrated luminosity, the square of the centre-of-mass energy, and the detection efficiency at the two c.m. energies, respectively. the number of background events for $J/\psi \to \bar\Lambda \Sigma^0$ is normalized to be {$669\pm120$}. It should be pointed out that there is no interference between the QED background and the $J/\psi$ resonance since this is a purely  electromagnetic process according to Ref.~\cite{Simone}.

\subsection{Signal Extraction\label{Signal}}
The signal yields are obtained from an extended unbinned maximum likelihood fit to the $\gamma\Lambda$ ($\gamma\bar{\Lambda}$) mass spectrum. The total probability density function (PDF) consists of a signal and various background contributions. The signal component is modeled as the MC simulated signal shape convolved with a Gaussian function to account for the difference in the mass resolution between data and MC simulation. The background components, $J/\psi\to\Sigma^0\bar\Sigma^0$, $J/\psi\to\Lambda\bar{\Lambda}$, and $J/\psi\to\gamma\Lambda\bar{\Lambda}~(\gamma\eta_c)$, as well as the reflection from signal conjugation decay mode, are described with the simulated shapes derived from the dedicated MC samples, while the magnitudes of different components are left free to account for the uncertainties of the branching fractions of these decays and other intermediate decays. The fit to the $M_{\gamma \Lambda}$/$M_{\gamma \bar \Lambda}$ spectrum, as displayed in Fig.~\ref{combineFits}, gives {$26260\pm181$} {$\bar{\Lambda}\Sigma^0$} events.


\subsection{Helicity Amplitude\label{HAmp}}
The structure of the six dimensional angular distribution is determined by global parameters $\bm{\omega}=(\alpha_{J/\psi}, \Delta\Phi, \alpha_\gamma, \alpha_{\Lambda}, \alpha_{\bar\Lambda})$ independent of the $\Sigma^0$ scattering angle, $\theta_{\Sigma^0}$, and is written in a modular form as
\begin{eqnarray}
  \begin{aligned}
    &\mathcal{W}\left(\bm{\xi}; \bm{\omega}\right)=\sum_{\mu,\nu=0}^{3} \sum_{\mu'=0}^{3} C_{\mu\nu} a_{\mu\mu'}^{\Sigma^0} a_{\mu'0}^{\Lambda} a_{\nu0}^{\bar\Lambda} \, ,
    \label{helicityequation}
  \end{aligned}
\end{eqnarray}
where the $C_{\mu\nu}(\theta;\alpha_{J/\psi}, \Delta\Phi)$ is a $4\times4$ spin density matrix, describing the spin configuration of the entangled hyperon-antihyperon pair.  The matrix elements are expressed as
\begin{eqnarray}
   \footnotesize
  \begin{aligned}
    &C_{\mu\nu}=(1+\alpha_{J/\psi} \cos ^{2} \theta)\begin{pmatrix}
      1  &  0  &  P_{y}  &  0  \\
      0  &  C_{xx}  &  0  &  C_{xz}  \\
      -P_{y}  &  0  &  C_{yy}  &  0  \\
      0  &  -C_{xz}  &  0  &  C_{zz}
      \end{pmatrix},
    \label{Cij}
  \end{aligned}
\end{eqnarray}
where $P_{y}$ governs the polarization of the $\Sigma^0$ and $C_{ij}$ characterizes its spin correlations. Both $P_{y}$ and $C_{ij}$ can be written  in terms  of $\sin\Delta\Phi$ or  $\cos\Delta\Phi$ as
\begin{equation}
  \begin{aligned}
   P_y=f(\theta) \sin\Delta\Phi, C_{xz}=f(\theta) \cos\Delta\Phi,
   \label{py_cxz}
  \end{aligned}
\end{equation}
where $f(\theta)$, a  common function dependent on the $\Sigma^0~(\bar\Sigma^0)$ direction angle $\theta$, is expressed as
\begin{equation}
  \begin{aligned}
    f(\theta)=\frac{\sqrt{1-\alpha_{J/\psi}^{2}} \sin\theta \cos\theta}{1+\alpha_{J/\psi} \cos^{2}\theta}.
  \end{aligned}
\end{equation}
The matrices $a_{\mu\nu}^{Y}$ in Eq.~\eqref{helicityequation} represent the propagation of the spin density matrices in the sequential decays. The full expressions of $C_{\mu\nu}$ and $a_{\mu\nu}^{Y}$ are given in Ref.~\cite{Perotti:2018wxm}.

\subsection{Global Fit of Parameters\label{Fit}}
A non-zero phase angle difference $\Delta\Phi$ indicates the transverse hyperon polarization, which allows us to measure these parameters at the same time. A simultaneous fit is performed to the two conjugate channels, $J/\psi\to\bar\Lambda\Sigma^0$ and $J/\psi\to\Lambda\bar\Sigma^0$. The likelihood function constructed from the probability density function for an event characterized by $\bm{\xi}_{i}$ is
\begin{eqnarray} \begin{aligned}
  \mathcal{L}=\prod_{i=1}^{N} \mathcal{P}\left(\bm{\xi}_{i}; \bm{\omega}\right)=\prod_{i=1}^{N} \frac{\mathcal{W}\left(\bm{\xi}_{i}; \bm{\omega}\right) \epsilon\left(\bm{\xi}_{i}\right)}{\mathcal{N}\left(\bm{\omega}\right)},
  \label{likelihood function}
\end{aligned}\end{eqnarray}
where $\epsilon\left(\bm{\xi}_{i}\right)$ is the detection efficiency, $N$ is {the number of} the surviving data events after all selection criteria, the normalization factor $\mathcal{N}\left(\bm{\omega}\right)=\int \mathcal{W}\left(\bm{\xi}; \bm{\omega}\right) \epsilon\left(\bm{\xi}\right) d\bm{\xi}$, with $\mathcal{W}\left(\bm{\xi}; \bm{\omega}\right)$ defined in Eq.~\eqref{helicityequation}, and $\mathcal{P}$ is the probability to produce event $i$ based on the measured parameters $\bm{\xi}_{i}$ and the set of observables $\bm{\omega}$.
Based on the likelihood function defined in Eq.~\eqref{likelihood function}, the objective function is written as
\begin{equation}
	S=-\ln \mathcal{L}_{\text {data}}^\text{I}-\ln \mathcal{L}_{\text {data}}^\text{II}+ \ln \mathcal{L}_{\text {bkg}}^\text{I}+ \ln \mathcal{L}_{\text {bkg}}^\text{II},
\end{equation}
where $\ln \mathcal{L}_{\text {data}}^\text{I,II}$ and $\ln \mathcal{L}_{\text {bkg}}^\text{I,II}$ are the likelihood functions for $J/\psi\to\bar\Lambda\Sigma^0$ and $J/\psi\to\Lambda\bar\Sigma^0$ and the background events from simulation, respectively. In order to optimize the free parameters ($\alpha_{J/\psi}$, $\Delta\Phi_1$ and $\Delta\Phi_2$) and minimize the objective function, the normalization factor $\mathcal{N}\left(\bm{\omega}\right)$ in Eq.~\eqref{likelihood function} is obtained by MC integral generated by phase space through all event selection criteria. To compare the fit with data, the moments directly related to helicity amplitude are defined as:
\begin{equation}\label{Item}
  \begin{aligned}
  T_{1}&=\sum_{i}^{N_k}\left(\cos ^{2} \theta~ n_{1, z}^{(i)}~ n_{2, z}^{(i)}-\sin ^{2} \theta~ n_{1, x}^{(i)}~ n_{2, x}^{(i)}\right), \\ 
  T_{2}&=\sum_{i}^{N_k} \cos \theta \sin \theta\left(n_{1, z}^{(i)}~ n_{2, x}^{(i)}-n_{1, x}^{(i)}~ n_{2, z}^{(i)}\right), \\ 
  T_{3}&=\sum_{i}^{N_k} \cos \theta \sin \theta~ n_{1, y}^{(i)}~, \\ 
  T_{4}&=\sum_{i}^{N_k} \cos \theta \sin \theta~ n_{2, y}^{(i)}~, \\ 
  T_{5}&=\sum_{i}^{N_k}\left(n_{1, z}^{(i)}~ n_{2, z}^{(i)}-\sin ^{2} \theta~ n_{1, y}^{(i)}~ n_{2, y}^{(i)}\right),
  \end{aligned}
\end{equation}
where $N_k$ is the number of events in the $k^{th}$ $\cos\theta$ bin and $\hat{\mathbf{n}}_{1}\left(\hat{\mathbf{n}}_{2}\right)$ is the unit vector in the direction of the nucleon (anti-nucleon) in the rest frame of $\Sigma^0$ ($\bar{\Lambda}$) for $J/\psi\to\bar\Lambda\Sigma^0$, as illustrated in Fig.~\ref{helicity}. The resulting $T_i$ distributions for data and the fit results are shown in Figs.~\ref{SigmaFitTi} and~\ref{SigmaBarFitTi}, and the difference between $T_3$ and $T_4$ results from the transverse polarization of $\Sigma^0~(\bar\Sigma^0)$, which allows the relative phase between $G_E$ and $G_M$ to be determined from the global fit of polarization with the modulus of the ratio between $G_E$ and $G_M$ obtained from $\alpha=\frac{s-4M^2_YR^2}{s+4M^2_YR^2}$.

\subsection{Systematic Uncertainty\label{Error}}
The uncertainties in the measurement of the form factors  are mainly from the $\Lambda, \bar\Lambda$ reconstruction, the 4C kinematic fit,  and the background estimation. For the $\Lambda, \bar\Lambda$ reconstruction, a correction to the MC efficiency is made. We also use the control sample of $J/\psi\to\bar{p}K^{+}\Lambda$ to obtain the efficiencies of the data and MC simulation in the $\Lambda$ and $\bar\Lambda$ reconstruction, and then correct the MC efficiencies by the observed data-MC efficiency differences. In order to reduce the impact of statistical fluctuations , the fit with the corrected MC sample is performed 400 times by varying the correction factor randomly within one standard deviation. The differences between the results with and without correction are taken as the systematic uncertainties. For the 4C kinematic fit, the MC sample in the polarization fit is altered by changing the helix parameters of charged tracks, and the same fit procedure is performed to the same data sample. The relative differences of the fit results are assigned as the uncertainties. The systematic uncertainty arising from the background estimate for each background source is assigned by varying the normalization factor by one standard deviation, the maximum change of the result is assigned as the associated systematic uncertainty. The total systematic uncertainty due to the background estimate is obtained by adding all effects of various background sources in quadrature. The uncertainties due to the $\alpha_{\Lambda, \bar\Lambda}$ are estimated by varying the quoted value from Ref.~\cite{BESIII:2022qax} within one standard deviation. The systematic uncertainties for the polarization measurement, as discussed above, are listed in Table~\ref{PolarTotalerr}.

\begin{table}[htp]
  \begin{center}
    \caption{Relative systematic uncertainties for the parameters {measurement} (\%).} 
    \vspace{0.50cm}
    \renewcommand\arraystretch{1.5}
    \setlength{\tabcolsep}{4mm}{
      \begin{tabular}{c|ccc}
        \hline\hline
        {Source}      & $\alpha_{J/\psi}$& $\Delta\Phi_1$ &     $\Delta\Phi_2$\\
        \hline
        $\Lambda/\bar\Lambda$ {reconstruction}  & {1.13}  & {0.   38} & {0.17}  \\
        Kinematic {fit}                         & {0.24} & {0.40}     & {0.23} \\
        Background {estimate}                   & {1.99}  & {0.   64} & {0.33}  \\
        $\alpha_{\Lambda/\bar\Lambda}$        & {0.00}  & {0.49}    & {0.09}  \\
        \hline
        Total                                 & {2.30} & {0.98} &     {0.45} \\
        \hline\hline
      \end{tabular}
    }\label{PolarTotalerr}
  \end{center}
\end{table}

\end{document}

%% file: authorlist_2022-08-01.tex
M.~Ablikim$^{1}$, M.~N.~Achasov$^{12,b}$, P.~Adlarson$^{72}$, M.~Albrecht$^{4}$, R.~Aliberti$^{33}$, A.~Amoroso$^{71A,71C}$, M.~R.~An$^{37}$, Q.~An$^{68,55}$, Y.~Bai$^{54}$, O.~Bakina$^{34}$
% ...
\\
\vspace{0.2cm}
(BESIII Collaboration)\\
\vspace{0.2cm} {\it
$^{1}$ Institute of High Energy Physics, Beijing 100049, People's Republic of China\\
% ...
$^{4}$ Bochum  Ruhr-University, D-44780 Bochum, Germany\\
% ...
$^{12}$ G.I. Budker Institute of Nuclear Physics SB RAS (BINP), Novosibirsk 630090, Russia\\
% ...
$^{33}$ Johannes Gutenberg University of Mainz, Johann-Joachim-Becher-Weg 45, D-55099 Mainz, Germany\\
$^{34}$ Joint Institute for Nuclear Research, 141980 Dubna, Moscow region, Russia\\
% ...
$^{37}$ Liaoning Normal University, Dalian 116029, People's Republic of China\\
% ...
$^{54}$ Southeast University, Nanjing 211100, People's Republic of China\\
$^{55}$ State Key Laboratory of Particle Detection and Electronics, Beijing 100049, Hefei 230026, People's Republic of China\\
% ...
$^{68}$ University of Science and Technology of China, Hefei 230026, People's Republic of China\\
% ...
$^{71}$ University of Turin and INFN, (A)University of Turin, I-10125, Turin, Italy; (B)University of Eastern Piedmont, I-15121, Alessandria, Italy; (C)INFN, I-10125, Turin, Italy\\
$^{72}$ Uppsala University, Box 516, SE-75120 Uppsala, Sweden\\
% ...
$^{b}$ Also at the Novosibirsk State University, Novosibirsk, 630090, Russia\\
% ...

}

%% file: draft_arxiv.bbl
\begin{thebibliography}{99}
  
  \bibitem{Hofstadter}
  R. W. McAllister and R. Hofstadter, \href{https://doi.org/10.1103/PhysRev.102.851}{{\it Phys. Rev.}~{\bf 102}, 851 (1956)}.
  
  \bibitem{EPJA51}
  V. Punjabi, C. F. Perdrisat, M. K. Jones, E. J. Brash and C. E. Carlson, \href{https://doi.org/10.1140/epja/i2015-15079-x}{{\it Eur. Phys. J. A}~{\bf 51}, 79 (1956)}.

  \bibitem{PRC85}
  A. J. R. Puckett {\it et al}. (Jefferson Lab Hall A Collaboration), \href{https://doi.org/10.1103/PhysRevC.85.045203}{{\it Phys. Rev. C}~{\bf 85}, 045203 (2012)}.

  \bibitem{PLB817}
  M. Ablikim {\it et al}. (BESIII Collaboration), \href{https://doi.org/10.1016/j.physletb.2021.136328}{{\it Phys. Lett. B}~{\bf 817}, 136328 (2021)}.

  \bibitem{PLB816}
  Y.~H.~Lin, H.~W.~Hammer and U.~Meißner, \href{https://doi.org/10.1016/j.physletb.2021.136254}{{\it Phys. Lett. B}~{\bf 816}, 136254 (2021)}.

  \bibitem{PRD104}
  A.~Mangoni, S.~Pacetti, and  E.~Tomasi-Gustafsson, \href{https://doi.org/10.1103/PhysRevD.104.116016}{{\it Phys. Rev. D}~{\bf 104}, 116016 (2021)}.

  \bibitem{Few-body}
  J.~Segovia, I.~C.~Cloet, C.~D.~Roberts, and S.~M.~Schmidt, \href{https://doi.org/10.1007/s00601-014-0907-2}{{\it Few Body Syst.}~{\bf 55}, 1185 (2014)}.

  \bibitem{Nuovo40}
  M. Ablikim {\it et al}. (BESIII Collaboration), \href{https://doi.org/10.1393/ncc/i2017-17028-3}{{\it Nuovo Cim. C}~{\bf 40}, 28 (2017)}.

  \bibitem{POS004}
  M. Ablikim {\it et al}. (BESIII Collaboration), \href{https://doi.org/10.22323/1.302.0004}{{\it PoS}~{\bf BORMIO2017}, 004 (2017)}.

  \bibitem{EWC192}
  M. Ablikim {\it et al}. (BESIII Collaboration), \href{https://doi.org/10.1051/epjconf/201819200023}{{\it EPJ Web Conf}~{\bf 192}, 00023 (2018)}.

  \bibitem{LambdaLambda}
  M. Ablikim {\it et al}. (BESIII Collaboration), \href{https://doi.org/10.1038/s41567-019-0494-8}{{\it Nature Phys.} \textbf{15}, 631 (2019)}.

  \bibitem{POS550}
  M. Ablikim {\it et al}. (BESIII Collaboration), \href{https://doi.org/10.22323/1.340.0550}{{\it PoS}~{\bf ICHEP2018}, 550 (2019)}.

  \bibitem{EWC199}
  M. Ablikim {\it et al}. (BESIII Collaboration), \href{https://doi.org/10.1051/epjconf/201919903009}{{\it EPJ Web Conf.}~{\bf 199}, 03009 (2019)}.

  \bibitem{POS121}
  M. Ablikim {\it et al}. (BESIII Collaboration), \href{https://doi.org/10.22323/1.346.0121}{{\it PoS}~{\bf SPIN2018}, 121 (2019)}.

  \bibitem{JPCS1137}
  M. Ablikim {\it et al}. (BESIII Collaboration), \href{https://doi.org/10.1088/1742-6596/1137/1/012010}{{\it J. Phys. Conf. Ser.}~{\bf 1137}, 012010 (2019)}.

  \bibitem{Nuovo42}
  M. Ablikim {\it et al}. (BESIII Collaboration), \href{https://doi.org/10.1393/ncc/i2019-19111-1}{{\it Nuovo Cim. C}~{\bf 42}, 111 (2019)}.

  \bibitem{ACP2130}
  M. Ablikim {\it et al}. (BESIII Collaboration), \href{https://doi.org/10.1063/1.5118406}{{\it AIP Conf. Proc.}~{\bf 2130}, 040009 (2019)}.

  \bibitem{JPCS1435}
  M. Ablikim {\it et al}. (BESIII Collaboration), \href{https://doi.org/10.1088/1742-6596/1435/1/012031}{{\it J. Phys. Conf. Ser.}~{\bf 1435}, 012031 (2020)}.

  \bibitem{PLB814}
  M. Ablikim {\it et al}. (BESIII Collaboration), \href{https://doi.org/10.1016/j.physletb.2021.136110}{{\it Phys. Lett. B}~{\bf 814}, 136110 (2021)}.

  \bibitem{PLB831}
  M. Ablikim {\it et al}. (BESIII Collaboration), \href{https://doi.org/10.1016/j.physletb.2022.137187}{{\it Phys. Lett. B}~{\bf 831}, 137187 (2022)}.

  \bibitem{PRD106}
  M. Ablikim {\it et al}. (BESIII Collaboration), \href{https://doi.org/10.1103/PhysRevD.106.L091101}{{\it Phys. Rev. D}~{\bf 106}, L091101 (2022)}.

  \bibitem{Lambdaform}
  M. Ablikim {\it et al}. (BESIII Collaboration), \href{https://doi.org/10.1103/PhysRevLett.123.122003}{{\it Phys. Rev. Lett.} \textbf{123}, 122003 (2019)}.

  \bibitem{Simone}
  R.~B.~Ferroli, A.~Mangoni, and S.~Pacetti, \href{https://doi.org/10.1140/epjc/s10052-020-08474-x}{{\it Eur. Phys. J. C} \textbf{80}, 903 (2020)}.

  \bibitem{Perotti:2020smi}
   E.~Perotti, \href{https://inspirehep.net/literature/1800840}{{\it Ph.D. thesis, Uppsala U.} (2020)}.

  \bibitem{BESIII:2021cxx}
  M. Ablikim {\it et al}. (BESIII Collaboration), \href{https://doi.org/10.1088/1674-1137/ac5c2e}{{\it Chin. Phys. C} \textbf{46}, 074001 (2022)}.

  \bibitem{Ablikim:2009aa}
  M. Ablikim {\it et al}. (BESIII Collaboration), \href{https://doi.org/10.1016/j.nima.2009.12.050}{{\it Nucl. Instrum. Meth. A} \textbf{614}, 345 (2010)}.

  \bibitem{Yu:IPAC2016-TUYA01}
  C.~H.~Yu {\it et al}., \href{https://doi.org/10.18429/JACoW-IPAC2016-TUYA01}{{\it Proceedings of IPAC2016, Busan, Korea,} 2016, doi:10.18429/JACoW-IPAC2016-TUYA01}.

  \bibitem{eemm}
  M. Ablikim {\it et al}. (BESIII Collaboration), \href{https://doi.org/10.1016/j.physletb.2019.03.001}{{\it Phys. Lett. B} \textbf{791}, 375 (2019)}.

  \bibitem{Im1-7}
  A. Z.~Dubnickova, S.~Dubnička, and M. P.~Rekalo, \href{https://doi.org/10.1007/BF02731012}{{\it Nuovo Cimento A} \textbf{109}, 241 (1996)}.

  \bibitem{Im2-8}
  S.~Pacetti, R. B.~Ferroli, and E.~Tomasi-Gustafsson, \href{https://doi.org/10.1016/j.physrep.2014.09.005}{{\it Phys. Rep.} \textbf{550}, 1 (2015)}.

  \bibitem{formfactor}
  G.~F\"aldt, and A.~Kupsc, \href{https://doi.org/10.1016/j.physletb.2017.06.011}{{\it Phys. Lett. B} \textbf{772}, 16 (2017)}.

  \bibitem{HyperonFFs}
  J.~Haidenbauer, U.~Meißner, and L. Y. Dai, \href{https://doi.org/10.1103/PhysRevD.103.014028}{{\it Phys. Rev. D}~{\bf 103}, 014028 (2021)}.

  \bibitem{BESIII:2022qax}
  M. Ablikim {\it et al}. (BESIII Collaboration), \href{https://doi.org/10.1103/PhysRevLett.129.131801}{{\it Phys. Rev. Lett.} \textbf{129}, 131801 (2022)}.

  \bibitem{gamma} 
  G.~F\"aldt, \href{https://doi.org/10.1103/PhysRevD.97.053002}{{\it Phys. Rev. D} \textbf {97}, 053002 (2018)}.

  \bibitem{whitebook}
  M. Ablikim {\it et al}. (BESIII Collaboration), \href{https://doi.org/10.1088/1674-1137/44/4/040001}{{\it Chin. Phys. C} \textbf{44}, 040001 (2022)}.

  \bibitem{ref1} 
  N.~Salone, P.~Adlarson, V.~Batozskaya, S.~Leupold, and J.~Tandean, \href{https://doi.org/10.1103/PhysRevD.105.116022}{{\it Phys. Rev. D} \textbf {105}, 116022 (2022)}.

  \bibitem{ref2} 
  G~Barucca {\it et al}., \href{https://doi.org/10.1140/epja/s10050-021-00386-y}{{\it Eur. Phys. J. A} \textbf{57}, 154 (2021)}.

  \bibitem{geant4}
  S.~Agostinelli {\it et al}., \href{https://doi.org/10.1016/S0168-9002(03)01368-8}{{\it Nucl. Instrum. Meth. A} \textbf{506}, 250 (2003)}.

  \bibitem{ref:kkmc}
  S.~Jadach, B.~F.~L.~Ward, and Z.~Was, \href{https://doi.org/10.1103/PhysRevD.63.113009}{{\it Phys. Rev. D} \textbf{63}, 113009 (2001)}; \href{https://doi.org/10.1016/S0010-4655(00)00048-5}{{\it Comput. Phys. Commun.} \textbf{130}, 260 (2000)}.

  \bibitem{ref:evtgen}
  D.~J.~Lange, \href{https://doi.org/10.1016/S0168-9002(01)00089-4}{{\it Nucl.\ Instrum.\ Meth.\ A} {\bf 462}, 152 (2001)}; R.~G.~Ping, \href{https://doi.org/10.1088/1674-1137/32/8/001}{{\it Chin. Phys. C} {\bf 32}, 599 (2008)}.

  \bibitem{PDG}
  P. A.~Zyla {\it et al}., \href{https://pdglive.lbl.gov/Viewer.action}{{\it Prog. Theor. Exp. Phys.} \textbf{2022}, 083C01 (2022)}.

  \bibitem{ref:lundcharm}
  J.~C.~Chen, G.~S.~Huang, X.~R.~Qi, D.~H.~Zhang, and Y.~S.~Zhu, \href{https://doi.org/10.1103/PhysRevD.62.034003}{{\it Phys.\ Rev.\ D} {\bf 62}, 034003 (2000)}; R.~L.~Yang, R.~G.~Ping, and H.~Chen, \href{https://doi.org/10.1088/0256-307X/31/6/061301}{{\it Chin.\ Phys.\ Lett.}\  {\bf 31}, 061301 (2014)}.

  \bibitem{conexc}
  R.~G.~Ping, \href{https://doi.org/10.1088/1674-1137/38/8/083001}{{\it Chin. Phys. C} \textbf{38}, 083001 (2014)}.

  \bibitem{Sigma0}
  B.~Zhong, R. G.~Ping, and Z. J.~Xiao, \href{https://iopscience.iop.org/article/10.1088/1674-1137/32/9/003}{{\it Chin. Phys. C} \textbf{32}, 692 (2008)}.

  \bibitem{VertexFit}
  M.~Xu {\it et al}., \href{https://doi.org/10.1088/1674-1137/33/6/005}{{\it Chin. Phys. C} \textbf{33}, 428 (2009)}.

  \bibitem{TopoAna}
  X. Y.~Zhou, S. X.~Du, G.~Li, and C. P.~Shen, \href{https://doi.org/10.1016/j.cpc.2020.107540}{{\it Comput. Phys. Commun.} {\bf 258}, 107540 (2021)}.

  \bibitem{QED}
  S. J.~Brodsky and G. R.~Farrar, \href{https://doi.org/10.1103/PhysRevLett.31.1153}{{\it Phys. Rev. Lett.} \textbf{31}, 1153 (1973)}.

  \bibitem{Perotti:2018wxm}
  E.~Perotti, G.~F\"aldt, A.~Kupsc, S.~Leupold, and J. J.~Song, \href{https://doi.org/10.1103/PhysRevD.99.056008}{{\it Phys. Rev. D}~{\bf 99}, 056008 (2019)}.

\end{thebibliography}
